\documentclass[12pt]{article}
\usepackage[textwidth=16cm,textheight=21cm]{geometry}

\usepackage{epsfig,multicol,multirow}
\usepackage{amsmath,subfigure,latexsym,amssymb}
\usepackage[table]{xcolor}

\usepackage{graphicx}
\usepackage{caption}
\usepackage[authoryear]{natbib}

\usepackage[hyphens]{url}

\usepackage[hidelinks]{hyperref}
\hypersetup{breaklinks=true}
\usepackage{ragged2e}

\newcommand{\be}{\begin{equation}}
\newcommand{\ee}{\end{equation}}

\newcommand{\bea}{\begin{eqnarray}}
\newcommand{\eea}{\end{eqnarray}}

\newcommand{\R}{{\kern+.25em\sf{R}\kern-.78em\sf{I} \kern+.78em\kern-.25em}}
\newcommand{\RR}{{\kern+.25em\sf{R}\kern-.6em\sf{I} \kern+.6em\kern-.25em}}
\newcommand{\N}{{\kern+.25em\sf{N}\kern-.78em\sf{I} \kern+.78em\kern-.25em}}
\newcommand{\C}{{\kern+.25em\sf{C}\kern-.50em\sf{I} \kern+.50em\kern-.25em}}

\usepackage{color}
\definecolor{col_red}{rgb}{1.0,0.0,0.0}

\makeatletter
\@addtoreset{equation}{section}
\makeatother

\begin{document}

\begin{flushright}
\today
\end{flushright}

\vspace*{3mm}

\begin{center}
{\large\bf Theoretical High Energy Physcis in Latin America}

\vspace*{6mm}

{\large\bf from 1990 to 2012: a Statistical Study} \\

\vspace*{1cm}

Gerardo Urrutia S\'{a}nchez$^{\rm \, a, b}$, Lilian Prado$^{\rm \, a}$ 
\vspace*{0.5mm} \\
and Wolfgang Bietenholz$^{\rm a}$
\ \\
\vspace*{5mm}

$^{\rm \, a}$  Instituto de Ciencias Nucleares \\
Universidad Nacional Aut\'{o}noma de M\'{e}xico \\
A.P.\ 70-543, C.P.\ 04510 Ciudad de M\'{e}xico, Mexico\\
\ \\ \vspace{-1mm}

$^{\rm \, b}$ Instituto de Astronom\'{\i}a\\
Universidad Nacional Aut\'{o}noma de M\'{e}xico \\
C.P.\ 04510 Ciudad de M\'{e}xico, Mexico\\

\end{center}

\vspace*{6mm}

\noindent

\noindent
We present a statistical overview of the publications in
theoretical high energy physics (HEP), which emerged in Latin 
America (LA) in the period from 1990 to 2012. Our study captures
the eight Latin American nations, which are dominant in this field
of research: Brazil, Mexico, Argentina, Chile, Colombia,
Venezuela, Uruguay and Cuba. As an intercontinental benchmark, 
we compare them with India, Canada, South Korea, Belgium and 
South Africa. We consider the productivity of research papers
in specialized high-impact journals, and the corresponding 
numbers of citations.
The goal is to document the efforts in LA to catch up with the
most wealthy countries, in a field of research without
direct practical benefits. The restriction to theoretical
HEP excludes large international collaborations, which enables
a fair evaluation of national achievements.
We further investigate how these records are correlated with
three socio-economic indices: the Gross Domestic Product (GDP),
the Human Development Index (HDI) and the Education Index (EI).
Despite some progress, there remains a backlog of LA compared to
the dominant countries, which cannot be explained solely by
economic deficiency. In general, a detailed correlation between the
socio-economic and scientific evolution is not obvious. \\
\ \\


\newpage

\tableofcontents

\section{Outline}

For centuries, the universities in Latin America (LA) were
entirely dedicated to higher education; 
only since the middle of the 20th century they also
act as research centers.
This was also the beginning of graduate programs in physics.
Half a century later, in the year 2000, LA had
a total population close to 500 millions, and about $10\,000$
active researchers in physics \citep{MoranLopez}.
This embraces some 800 active
researchers in High Energy Physics (HEP) --- including Ph.D.\
students --- with a majority of about 650 researchers
working in theoretical HEP \citep{Masperi}.
  
Our study sets in four decades after a LA research community
in physics (beyond individuals) was established. More precisely,
this work deals with research papers in theoretical HEP from LA,
in the period from 1990 and 2012. HEP is a specific field
of science which does not focus on immediate practical applications.
The motivations for this line of research is rather
cultural and pedagogical, whereas technological
benefits are only conceivable on a longer time-scale.

As a distinction from the previous bibliometric literature,
our restriction to theory excludes large international collaborations,
where national contributions are hard to quantify, and which could
therefore significantly distort the statistics of publications
and citations.
  
We present statistical data for the productivity of the
eight dominant countries regarding the number of publications,
the Impact Factor (IF) of the
journals where they appeared, and the citations that they received.
In order to interpret these data, we compare them to
five intercontinental countries, and we search for correlations
with the economic and social development.

Section 2 explains our selection criteria for the data that
enter our statistics. Section 3 presents the record for each country
under consideration, with respect to the number of publications,
the IF of the corresponding journals,
and the number of citations to these articles. Section 4 confronts
these results with three well-established socio-economic indices.
Section 5 is devoted to our conclusions, and an appendix gives
an explicit list of particularly top-cited and groundbreaking papers.

\section{Criteria for the statistical data}

This section explains how the entries to our statistics were
selected. We consider peer-reviewed publications\footnote{In
HEP also preprints play an important r\^{o}le,
more than in other fields, in particular due to the data bases
{\em arXiv} and {\em INSPIRE} (formerly SPIRES). For instance in
2015 a total of 10\,126 preprints were submitted to the arXiv;
6415 of them were later published in peer-reviewed journals,
unlike proceeding contributions (about $20\,\%$) and others
\citep{Brasil}. In INSPIRE-HEP we capture contributions
to hep-ph ($48\, \%$), hep-th ($40\, \%$) and hep-lat ($4\, \%$),
but not to hep-ex ($8\, \%$).}
that are documented in the {\em Web of Science}
--- formerly known as the {\em Web of Knowledge} ---
\citep{wok}, which is widely accepted as a
reliable data base of the scientific literature.
A paper is counted for the productivity of a country {\em if at
least one author indicates a working address}
{\em in that country.}
(We are not concerned with the authors' citizenships.)
Hence some publications count for several
countries, although relatively few papers emerge from international
collaborations within LA; a study of all scientific papers
\citep{Russell} found that this fraction was always below
$4\,\%$ from 1975 to 2004 (most frequent were collaborations
between Argentina and Brazil).

Our selection rule is sensible only as long as the number of authors
of an article is modest. Large international collaborations --- such
as the four LHC collaborations at CERN, with thousands of members 
--- may be very productive, but their papers would confuse the
interpretability of such a statistics: very few collaboration
members in one country could contribute ${\cal O}(100)$ papers
per year, but as a very small group among thousands of authors; that
could drastically overestimate the importance of their contribution.

Since the 1990s --- when experimental HEP gained new momentum
--- some LA countries
got strongly involved in ``Big Science'' projects (large international
experimental collaborations), which boosted the HEP community
and enhanced its overall HEP productivity
(to some extent, this process had already started in the 1980s
\citep{Masperi}).
The case of Mexico is analyzed in \cite{HEPmex,BigSci04},
which describe the participation in research centers like CERN,
DESY, SLAC, Fermilab and BNL as a style of science, which was
new in LA. The HEP challenges required a new working style,
which later propagated into other sciences \citep{Manga}.
\cite{BigSci04} observed that experimental arXiv entries
receive on average about 4.5 times more citation than theoretical
ones, but that theoretical papers have a longer
``citing life'', and that cross-citations
between theory and experiment are rare.
A number of papers by large collaborations are very highly cited,
and including them would clearly distort our statistics.
\cite{Manga} document this effect in the case of
Brazil: in the period 2010-14 its participation in large
collaborations yielded just $8.3\,\%$ of the HEP papers,
but $33.2\,\%$ of the citations. Hence this small fraction
strongly affects the citation record (and therefore also
the university rankings), if all HEP is combined.
\cite{enrich} reported a similar phenomenon in Mexico.
This confirms that experimental HEP
may well have a misguiding effect on statistical studies.
Theoretical HEP, however, is particularly appropriate for
a clean analysis of national achievements
(mathematics might be an alternative).

This motivates our restriction to papers in 
{\em theoretical} HEP, where the number of authors is usually
small, so one can assume each of them to have contributed significantly.
To the best of our knowledge, this is the first study of this kind.
As a quantitative rule, our statistics only includes papers signed
by {\em less than 20 authors.}

This cutoff tends to include (exclude) theoretical (experimental)
research papers.
Referring to HEP in general, \cite{credit} points out that
there is more ``hyper-authorship'' (inclusion of passive authors)
than in other disciplines. This obviously concerns experimental
physics, which again justifies our focus on
theory.\footnote{\cite{credit2}, as well as \cite{credit} further
observed that in about 3/4 of the HEP papers, the authors appear
in alphabetic order. This quota is higher than in other disciplines,
and it shows that in our study a consideration of first authors
is not motivated.}

We also had to be selective with the journals where the articles
contribute to our statistics. Unfortunately we had to {\em exclude
interdisciplinary journals,} like Physics Review Letters,
Nature and Science, although they published some particularly
important works in theoretical HEP. However, in the framework of
our statistical study it would have been practically
impossible to select the papers in these journals which refer
to theoretical HEP.\footnote{\cite{Mele06} observed in 2006
that a total of about 6000 HEP papers are published annually
(including experimental HEP), $83 \, \%$ of them in just
six journals. Five of them are specialized on HEP and therefore
included in our Table \ref{jtab} (Phys.\ Rev.\ D, Phys.\ Lett.\ B,
Nucl.\ Phys.\ B, J. High Energy Phys., Eur.\ Phys.\ J.\ C),
whereas Phys.\ Rev.\ Lett.\ is broadband, as we mentioned above.}
Therefore our data taking is limited to {\em specialized journals,} which 
are exclusively {\em devoted to HEP,} such that all their articles (that
refer to theory) can be considered contributions to the field that we 
consider.

Moreover, in order to restrict the data to articles of scientific
importance, we only considered {\em high-impact journals.} In this
respect, our criterion was a {\em 2-years IF $>1$} in the year 
2012, according to the \cite{JCR}.\footnote{To compute the IF
of some year $Y$, one considers all articles published in one
journal during the years $Y-1$ and $Y-2$.
The IF is the average number of citations that these articles
received in the course of the year $Y$ \citep{JCR}.}

The application of these selection criteria led to a set of
14 international journals, which do provide entries to our
statistics. They are listed in Table \ref{jtab}, along with their IF;
before 2010 it is averaged over periods of 5 years. Note that
five of these journals came to existence after 1990,\footnote{This
concerns J.\ Cosmol.\ Astropart.\ Phys.\ (since 2003), 
J.\ High Energy Phys.\ (since 1997), 
Eur.\ Phys.\ J.\ C (since 1998),
Astropart.\ Phys.\ (since 1992) and
Adv.\ High Energy Phys.\ (since 2010).\label{fnjdat}}
which explains the empty slots in Table \ref{jtab}.
\begin{table}
\centering
{\small
\begin{tabular}{|l||c|c|c|c|c|}
\hline 
 & 1990--94 & 1995--9 & 2000--04 & 2005--09 & 2012 \\
\hline
\hline
J.\ Cosmol.\ Astropart.\ Phys. & --- & --- & 7.914 & 6.374 & 6.036 \\
\hline 
J.\ High Energy Phys. & --- & --- & 6.454 & 5.678 & 5.618 \\
\hline 
J.\ Phys.\ G 
& 2.178 & 1.277 & 1.348 & 2.966 & 5.326 \\
\hline 
Eur.\ Phys.\ J.\ C & --- & --- & 4.766 & 3.453 & 5.247 \\
\hline
Astropart.\ Phys. & --- & --- & 3.924 & 3.783 & 4.777 \\
\hline
Phys.\ Rev.\ D & 2.734 & 3.702 & 4.462 & 4.883 & 4.691 \\
\hline
Phys.\ Lett.\ B & 3.174 & 3.723 & 4.314 & 4.291 & 4.569 \\
\hline
Nucl.\ Phys.\ B & 4.578 & 3.311 & 5.395 & 4.771 & 4.327 \\
\hline
Class.\ Quantum Gravity & 1.492 & 1.790 & 2.262 & 2.924 & 3.562 \\
\hline
Adv.\ High Energy Phys. & --- & --- & --- & --- & 3.500 \\
\hline
Prog.\ Part.\ Nucl.\ Phys. & 2.060 & 2.119 & 2.354 & 3.699 & 2.257 \\
\hline
Int.\ J.\ Mod.\ Phys.\ A & 1.411 & 1.400 & 1.198 & 1.014 & 1.127 \\
\hline
Mod.\ Phys.\ Lett.\ A & 1.306 & 1.070 & 1.251 & 1.335 & 1.110 \\
\hline
Int.\ J.\ Theor.\ Phys.\ & 0.370 & 0.438 & 0.556 & 0.530 & 1.086 \\
\hline
\end{tabular}
}
\caption{List of the journals specialized on HEP,
with their 2-years IF averaged over periods of 5 years, and in
2012. The latter is above 1 in all cases, which was our
criterion for a high-impact journal, where the articles enter
our statistics.}
\label{jtab}
\vspace*{-3mm}
\end{table}

Based on these criteria, our study captures theoretical research
papers in {\em particle physics} and {\em quantum field theory,} 
and --- following general conventions \citep{inspire} --- also 
in {\em cosmology} and {\em gravity} (although the term ``high 
energy'' might be arguable in those cases). On the other hand,
for instance the extensive literature on condensed matter physics
and optics is not included.\footnote{According to \cite{Russell},
between $10$ and $15\, \%$ of the LA physics papers deal with
particles and fields.}
As we mentioned in Section 1, we consider subjects which are rather
far from practical applications, say in technology or industry, at
least on a short time-scale.
Hence these fields of research can be considered as some kind of ``luxury''.
On the other hand, they are relatively cheap, compared to 
experimental science.
These characteristics should be kept in mind in Section 4, when
we are going to compare the national productivities with indices 
from economy and development.

Our search strategy in the Web of Science specified a country,
publications in one of the journals in Table \ref{jtab}, research
in ``Physics Particles Fields'', and the restriction to less than
20 authors. This excludes most experimental works, but some of them
adhere to this cutoff. To exclude them we eliminated the research
areas ``instrumentation'' and ``spectroscopy'', as well as some
well-known experimental collaborations. Then we searched
for specific terms in title, keywords and abstract, which
could hint at experimental physics, like ``measurement'', ``beam'',
``detection'',  ``instrumentation'',  ``spectrum'', ``collision''
and ``eV'' (the latter is also sensitive to ``MeV'', ``GeV'' etc.).
If such a suspicious term was detected, a look at title and abstract
was necessary to decide whether the paper was experimental or theoretical.

To summarize: {\em articles enter our statistics if they were
published in one of the journals on Table \ref{jtab}, if they
refer to theoretical physics, and if the number of authors is
below 20. They count for a country if at least one author gives
a working address there.} 
Citations to these articles count in any paper which appears in 
the Web of Science, until (and including) the year 2012.

\section{Ranking by nations}

\subsection{Overview}

The left-hand side of Table \ref{pubtab} gives the total 
number of publications --- according to the criteria 
specified in Section 2 --- that emerged
in the eight dominant LA countries. They can be divided into a
{\em leading group: Brazil, Mexico, Argentina} and {\em Chile,} and a
{\em sub-leading group,} consisting of {\em Colombia, Venezuela,
Uruguay} and {\em Cuba.} All other countries in LA have a production
rate below one typical active researcher, who publishes about
two papers per year; this motivates our cutoff at this point.

The corresponding numbers for the five intercontinental countries
under consideration are given on the right-hand side of Table 
\ref{pubtab}.\footnote{This table also contains the official 
  national acronyms, to be used in the following.}
We see that their productivity is in the magnitude 
of the leading group in LA.

Regarding science and technology in general,
a comparison between the leaders in LA and among the
intercontinental countries in Table \ref{pubtab}, Brazil and
India, is discussed by \cite{Sikka}, which stresses certain
similarities. A newer study of this kind, \cite{impact}, considers
nine developing countries, including BRA, MEX, ARG, CHL, IND
and ZAF. It applies methods (like the illustration with ``radar
charts'') of the classical global analysis by \cite{King}.
Below we are going to comment on key observations by \cite{impact}.

From a global perspective, \cite{Mele09} reports that
{\it e.g.}\ in the years 2005/6 the countries included in our
studies produced the following fractions of the HEP papers
worldwide: CAN $2.8\, \%$, BRA $2.7\, \%$, IND $2.7\, \%$,
KOR $1.8\, \%$, MEX $0.8\, \%$, BEL $0.7\, \%$, ARG $0.6\, \%$,
CHL $0.6\, \%$, while the rest is below $0.4\, \%$. This
hierarchy is similar to Table \ref{pubtab}; the most notable
difference is that Canada and Brazil catch up with India.
Explanations could be the inclusion of experimental HEP,
and that co-authorship is counted on a {\it pro-rata} basis
(assigning fraction of a country).

These results can be compared to the ranking by \cite{MoranLopez},
which divides the LA countries into four groups,
based on the number of Ph.Ds in physics and ``overall scientific
output'':\\
Group 1: Brazil, Mexico, Argentina.\
Group 2: Cuba, Chile, Venezuela, Colombia.\\
Group 3: includes 11 countries, among them Uruguay.\
Group 4: rest, negligible in physics.\\
This is very similar to our groups Table \ref{pubtab},
but there Chile and Uruguay move one level up. 
  
\begin{figure}[h!]
\begin{center}
\includegraphics[width=0.555\textwidth,angle=0]{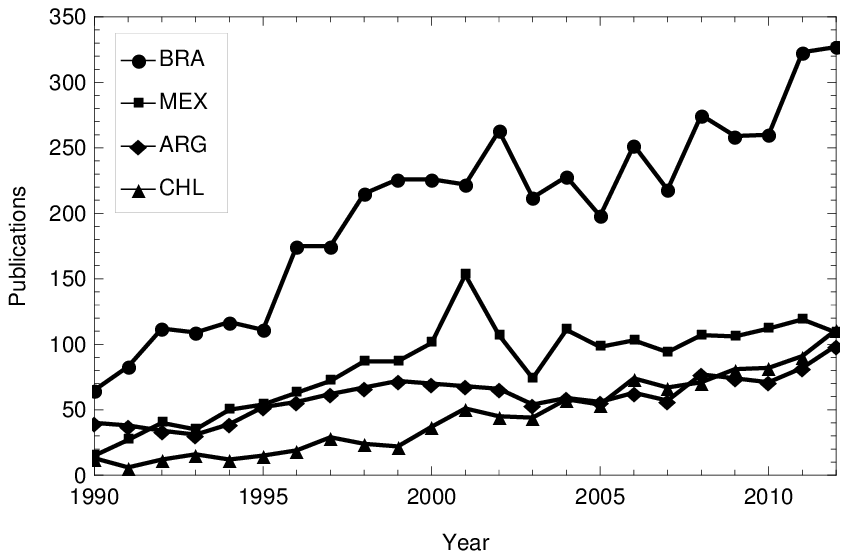}
\includegraphics[width=0.555\textwidth,angle=0]{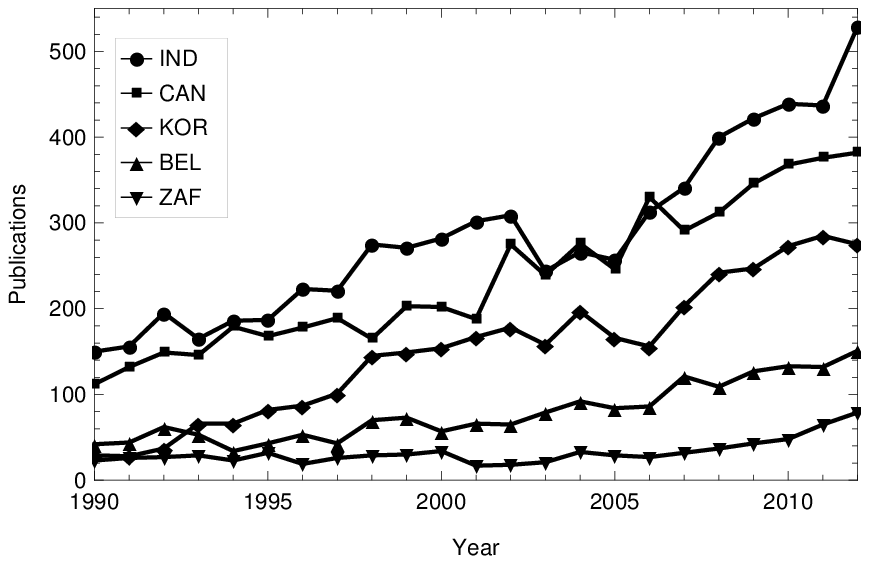}
\includegraphics[width=0.545\textwidth,angle=0]{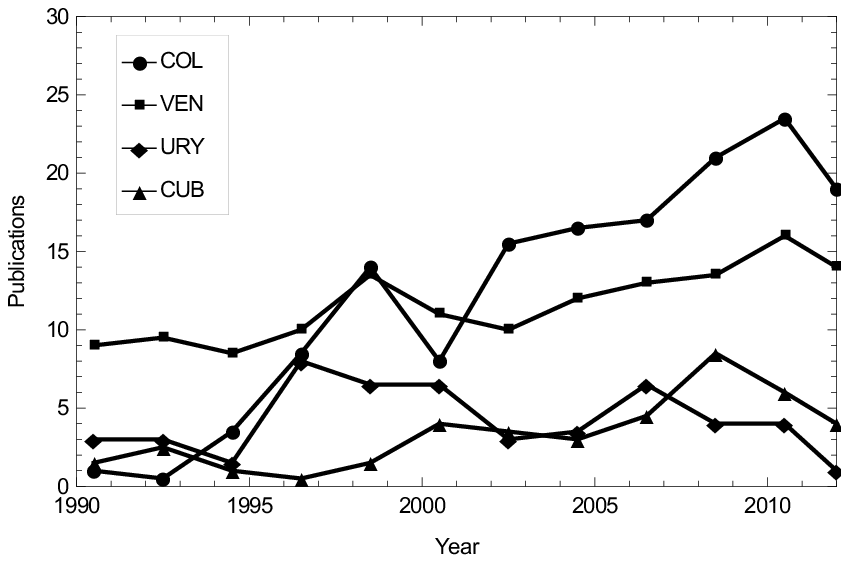}
\caption{Time-line of the publication rate, from 1990 to
  2012, for the countries under consideration
  (the acronyms are given in Table \ref{pubtab}).
  The upper two plots display the number of publications each year.
  In the plot below 2-years averages were taken (plus the value of
  2012) in order to smoothen large relative fluctuations.}
\vspace*{-6mm}
\label{pubtimeline}
\end{center}
\end{figure}

The time-lines which led to Table \ref{pubtab}, {\it i.e.}\ the
production in these countries during the period from 1990 to 2012,
is illustrated in Figure \ref{pubtimeline} 
(for the detailed numbers we refer to \cite{tesis}).
The general trend is that the productivity increases with time
(up to short term fluctuations). For some --- though not all ---
countries we observe a dip, in the period
(roughly) between 2001 and 2004, to be interpreted in Subsection 4.3.

\begin{table}
\centering
\begin{tabular}{|l|c|r|r|c|l|c|r|r|}
\hline
 & & publi- & & & & & publi- & \\
 & & cations & rwf & & & & cations & rwf \\
\hline
\hline 
Brazil & BRA & 4650 & 3.68 & ~~ & India & IND & 6570 & 3.39 \\
\hline 
Mexico & MEX & 1924 & 3.33 & & Canada & CAN & 5452 & 4.03 \\
\hline 
Argentina & ARG & 1387 & 3.73 & & South Korea & KOR & 3491 & 4.23 \\
\hline 
Chile & CHL & 1034 & 4.23 & & Belgium & BEL & 1819 & 4.08 \\
\hline 
 & &  & & & South Africa & ZAF & 747 & 3.51 \\
\hline 
\hline
Colombia & COL & 277 & 3.88 & & & & & \\
\hline
Venezuela & VEN & 266 & 3.48 & & & & & \\
\hline
Uruguay & URY & 100 & 3.63 & & & & & \\
\hline
Cuba & CUB & 77 & 3.72 & & & & & \\
\hline
\end{tabular}
\caption{The total number of publications, which fulfil the 
criteria of Section 2, over the entire period from 1990 to 
2012. The gap in the left columns (which refers to LA) distinguishes 
the leading from the sub-leading group among these eight countries.
The intercontinental countries (right columns) are comparable to
the leading group in LA. The column ``rwf'' shows the average 
re-weighting factor by the journal IF, as described in Subsection 
3.2. We also display the official acronym for 
each country, to be used below.}
\label{pubtab}
\end{table}

\subsection{Re-weighting by the journal Impact Factor}

\begin{figure}[h!]
\begin{center}
\includegraphics[width=0.8\textwidth,angle=0]{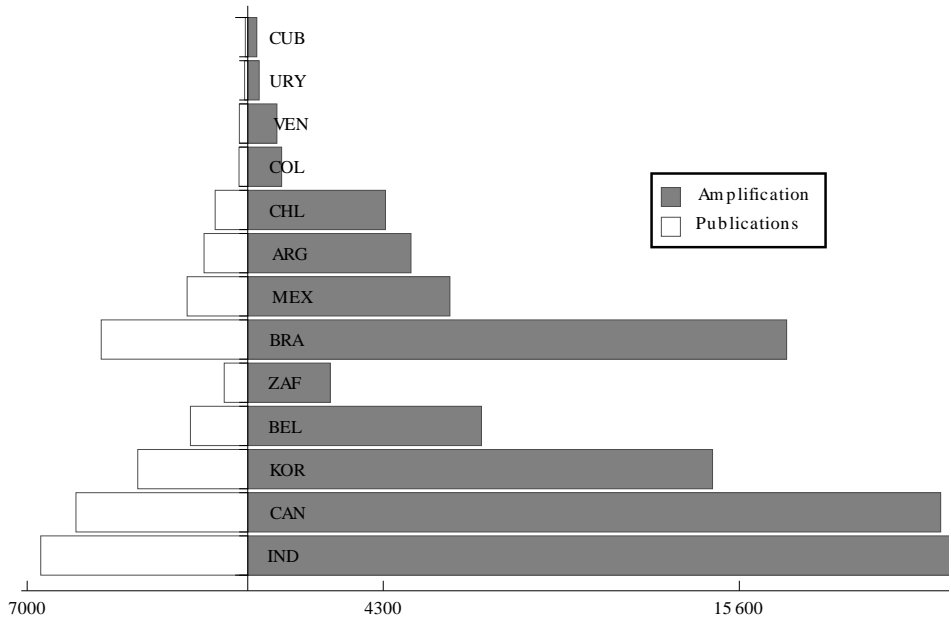}
\vspace*{-4mm}
\caption{Histograms of the total number of publications per country
(to the left), and of the IF-re-weighted record (to the right).}
\vspace*{-5mm}
\label{IFamp}
\end{center}
\end{figure}

A possible criticism of the overview in Subsection 3.1 is that
all articles are counted equally, if they refer to theoretical HEP 
and they were published in any of the journals of Table \ref{jtab}, which 
all had an IF $>1$ in the year 2012. However, in the considered period
the IFs of these journals vary from 0.370 to 7.914, so one could
argue that these articles should be counted with different weights.
Figure \ref{IFamp} shows the outcome of such a re-weighting, where 
each paper is multiplied with the IF of its journal in that year.

Regarding LA, the hierarchy among the eight dominant countries
does not change, but the dominance of Brazil is further enhanced.
In particular, the structure of a leading and a sub-leading group,
composed of four countries each, persists. 
This distinction is not visible, however, if we just consider the
mean re-weighting factor for each country, which is given
in the columns ``rwf'' of Table \ref{pubtab}. If we include
the intercontinental countries, we see that Chile, South Korea,
Belgium and Canada tend to publish in high-impact journals
(rwf $>4$), whereas the papers from South Africa, Venezuela, 
India and Mexico appeared in journals with relatively low IF.

\begin{figure}[h!]
\begin{center}
\includegraphics[width=0.7\textwidth,angle=270]{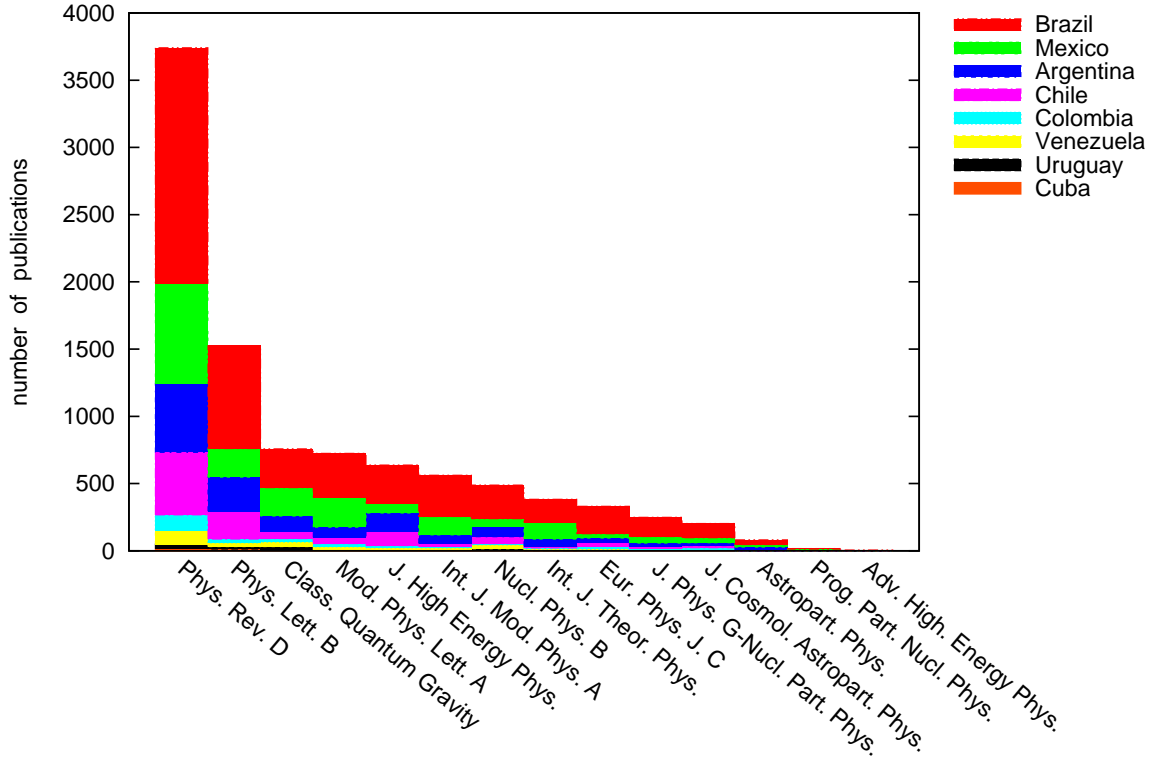}
\vspace*{-4mm}
\caption{Histogram for the number of publications from LA
in the journals of Table \ref{jtab}. 
In each bin, the rectangle is divided into the contributions
by the eight countries under consideration. Phys.\ Rev.\ D
clearly dominates, followed by Phys.\ Lett.\ B.}
\vspace*{-5mm}
\label{Jhisto}
\end{center}
\end{figure}

Figure \ref{Jhisto} shows how many papers from LA were 
published in each of the 14 journals of Table \ref{jtab}. 
There is a clear dominance by Phys.\ Rev.\ D in all countries,
usually followed by Phys.\ Lett.\ B (exceptions are Mexico and
Venezuela, where Mod.\ Phys.\ Lett.\ A and Class.\ Quantum Gravity
is second, respectively).
We recall, however, that five of these 
journals were founded after 1990, cf.\ footnote \ref{fnjdat}.
For detailed numbers and the annual time evolution, we refer 
again to \cite{tesis}, since a comparative study 
of the importance of various journals is not our goal.

\subsection{Total number of citations}

Yet another point of view is that one should directly re-weight
each article with the number of citations to it, rather than 
using the IF of the journal. This amounts to a statistics of 
all citations received (at any time from 1990 to 2012) by all 
papers of one country. This sum,
over all 23 years, is shown in Table \ref{cittab}.
Within LA the striking dominance of Brazil persists
(in particular the citation ratio Brazil/Mexico, $2.45$, is
very similar to the publication ratio, $2.42$), while Argentina almost 
catches up with Mexico, and also Chile is moving closer.

\begin{table}
\centering
\begin{tabular}{|l|r|r|c|l|r|r|}
\hline
 & cita-~ & citations~~ & & & cita-~ & citations~~ \\
 & tions~ & per article & & & tions~ & per article \\
\hline
\hline 
 & & & ~~ & Canada & 105189 & 19.3 \\
\hline 
 & & & & India & 89100 & 13.6 \\
\hline 
Brazil & 53452 & 11.5 & ~~ & South Korea & 50894 & 14.6 \\
\hline 
Mexico & 21797 & 11.3 & & Belgium & 32951 & 18.1 \\
\hline 
Argentina & 19164 & 13.8 & & & &  \\
\hline 
Chile & 15284 & 14.8 & & & & \\
\hline 
 & &  & & South Africa & 10455 & 14.0 \\
\hline 
\hline
Venezuela & 4018 & 15.1 & & & &  \\
\hline
Colombia & 3476 & 12.5 & & & & \\
\hline
Uruguay & 1628 & 16.3 & & & & \\
\hline
Cuba & 713 & 9.3 & & & & \\
\hline
\end{tabular}
\caption{The total number of citations received until 2012.
  We also display the mean number of citations per article.
  In that regard, Uruguay, Venezuela and Chile achieved the best
  values in LA. Note, however, that Uruguay's value is to a
  significant part due to one single publication, quoted as
  Ref.\ [2] in Appendix A.1; otherwise it would have just 12.3
  citations per article.}
\label{cittab}
\end{table}

We observe a marked distinction between these four leading nations 
and the sub-leading group, also with respect to the citations. 
Within the latter group, we notice that Venezuela is slightly 
superior to Colombia, in contrast to the hierarchy of Table \ref{pubtab}.

Earlier statistics of the citations of LA articles from all sciences
are given by \cite{Kraus} and \cite{Osa}. The latter refers to
the period 1981-93, and arrives at the same hierarchy as the leading
group as in Table \ref{cittab} (which also coincides with
Table \ref{pubtab}): BRA, MEX, ARG, CHL.

Compared to Table \ref{pubtab}, the intercontinental countries appear 
superior in this statistics. In particular, Canada achieved an
impressive number of more than $10^{5}$ citations. The trend is that
papers from LA are somewhat less quoted than the publications from
these five countries; the overall average citation number per article
in LA (intercontinental) is 12.30 (15.96).

\begin{figure}[h!]
\begin{center}
  \hspace*{-6mm}
\includegraphics[width=0.55\textwidth,angle=0]{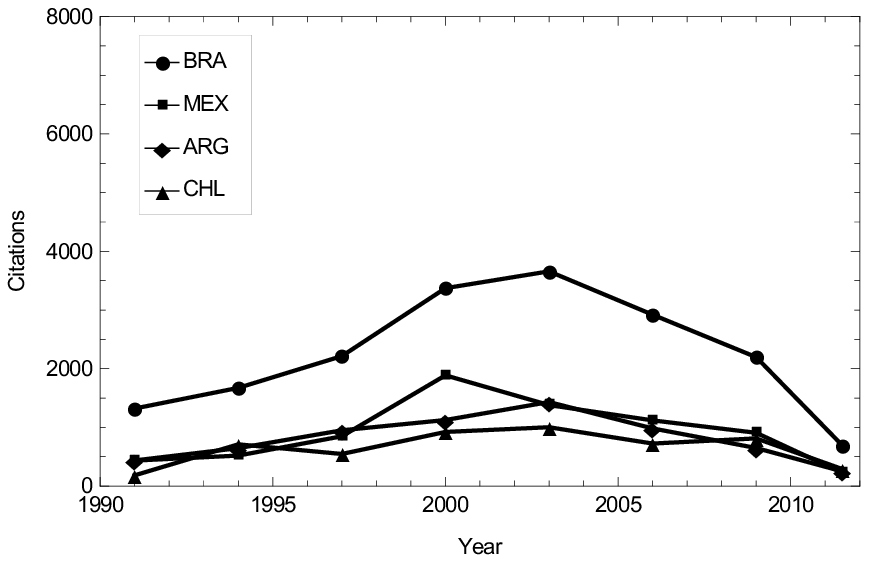}\\
\includegraphics[width=0.555\textwidth,angle=0]{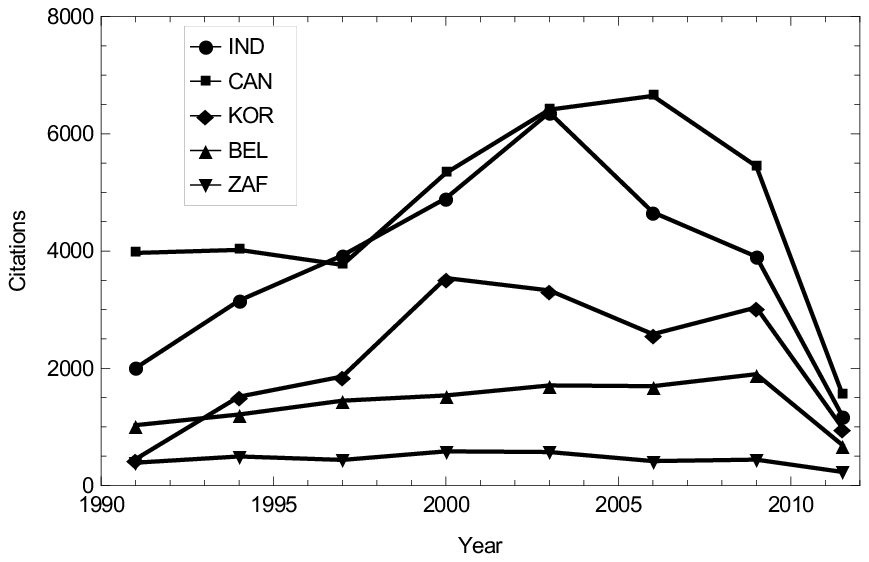}\\
\includegraphics[width=0.55\textwidth,angle=0]{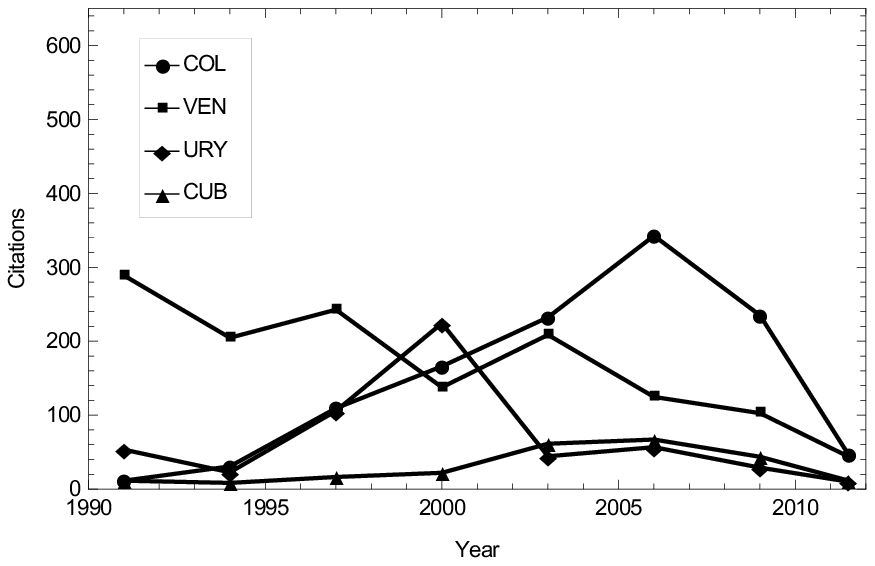}
\vspace*{-1mm}
\caption{Time-lines of the citations received, from 1990 to 2012,
  for the countries under consideration. We show 3-years averages
  (plus the average of 2011 and 2012).}
\vspace*{-6mm}
\label{citimeline}
\end{center}
\end{figure}

In order to make this point more explicit, Table \ref{cittab} also 
displays the mean number of citations per article for each country.
The time-lines of citations are shown in Figure 
\ref{citimeline}. In general, the world-wide publication output 
tends to rise with time, which explains an increasing 
number of citations until the end of the 20th century. The decrease 
after 2005 is simply due to the fact that newer papers
had little time to be cited (before 2013).

The results of this section can be compared to the data on the
webpage \cite{scimagojr}, which displays a ranking for LA
(or other regions) in Nuclear Physics and HEP, since 1996.
That ranking is based on the national H-index \citep{Hirsch}.
If we consult it for single years, we see that our dominant group
is practically the same (as a difference we note that until 1999
ARG was second, before MEX). Among the following four countries,
the annual ranking sometimes contains Ecuador, Peru and Puerto
Rico, and otherwise countries which are present in our study.
The summary of the period 1996-2016 is identical to our Table
\ref{pubtab} up to position 6, followed by Cuba, Ecuador, Puerto
Rico, Peru and Uruguay. These slight differences are apparently
due to the inclusion of experimental HEP, and nuclear
physics.\footnote{Our study does include nuclear physics at the
  fundamental level of Quantum Chromodynamics, but not necessarily
  the (more traditional) effective approaches to nuclei, with
  external potentials etc.}

In Appendix A.1 we add a list of the most cited LA papers
which enter our statistics.

\subsection{Publications {\it per capita}}

Depending on the aspect of interest, one should rather
refer to the scientific productivity of different countries
{\em relative to their population.} In fact, the populations of
the eight LA countries in our tables varies by orders of 
magnitude.\footnote{From 1990 to 2012, the population (in millions)
evolved as follows: Brazil 149 to 198, Mexico 83 to 115, 
Colombia 34 to 47, Argentina 32 to 41, Venezuela 19 to 30, 
Chile 13 to 17, Cuba 10.6 to 11.3, Uruguay 3.1 to 3.4.}
Table \ref{pubpoptab} shows, country by country, 
the total number of papers, divided by the population (in millions) 
of that country, in the year when each paper was published.
This table also includes the total number of citations per
million of inhabitants in 2012.

\begin{table}
\centering
\begin{tabular}{|l|r|r||l|r|r|}
\hline
 & publications & citations & & publications & citations \\
 &  per million & per million & 
 &  per million & per million \\
 & inhabitants & inhabitants & 
 & inhabitants & inhabitants \\
\hline
\hline 
 & & & Belgium & 173.50 & 2969.9 \\
\hline 
 & & & Canada & 172.19 & 3020.3 \\
\hline
Chile & 63.90 & 878.2 & South Korea & ~72.91 & 1017.7 \\
\hline 
Argentina & 37.60 & 467.1 & & & \\
\hline 
Uruguay & 30.54 & 481.5 & & & \\
\hline 
Brazil & 25.89 & 269.5 & & & \\
\hline 
Mexico & 18.93 & 189.8 & South Africa & 16.49 & 204.2 \\
\hline 
Venezuela & 10.80 & 136.1 & & & \\
\hline
Cuba & ~6.88 & 63.3 & & & \\
\hline
Colombia & ~6.49 & 74.6 & India & 6.18 & 72.8 \\
\hline
\end{tabular}
\caption{The total number of publications, and citations, divided 
by the population of each country, in millions of inhabitants.}
\label{pubpoptab}
\end{table}

We observed before that re-weighting by the IF, or by the citation 
numbers, hardly affects the hierarchy. The statistics
{\it per capita,} however, does alter the picture significantly.
Here the leader turns out to be Chile, followed by Argentina, 
and Uruguay rises from the sub-leading group to the third position.
On the other hand, Brazil and Mexico lose three positions each, and
Colombia slips down to the last position in this list for LA.

This matches the observation of \cite{impact}
that Chile has a particularly high number of publications and
citations per researcher in  science and technology.
This observation is also consistent with data available
from the public webpage \cite{RICYT}. As an
interpretation, \cite{impact} refer to Chile's excellent
collaboration with Europe, in particular related to the
European Southern Observatory for ground-based astronomy,
created by 16 nations in 1962, and located in northern Chile.
However, our observation of an analogous trend in theoretical
HEP suggests that this observatory alone is not sufficient
to explain Chile's extraordinary scientific efficiency.

From the intercontinental perspective, we now see an
overwhelming record of the First World countries (Canada and
Belgium), which might be expected in a field of research
without immediate applications. South Korea is still ahead of
all LA countries, whereas India --- the leader in Table \ref{pubtab}
--- is now below any of the eight LA nations in our tables.

The perspective of this subsection will be relevant 
for considerations in Section 4, in particular 
for the search of correlations with human development and education.

\section{Correlations with socio-economic indices}

\subsection{Gross Domestic Product (GDP)}

The GDP is widely used as a measure for the economic performance
of a country \citep{GDP}. It represents the monetary value --- in US 
dollars --- of all final goods produced, and services provided, 
during one year.\footnote{This is not based on the
  currency exchange rate, but on the actual values of goods,
  {\it i.e.}\ it refers to {\em purchasing power parity.}}
Three quantities can be used to estimate the GDP; theoretically
the results should coincide.

\begin{itemize}

\item Total value of the domestic production and services,
minus intermediate consumption.

\item Sum of the incomes of all residents and enterprises.
(Unpaid work is not captured.)

\item Sum of all expenses for purchasing final goods and services, 
which emerge in the country under consideration.

\end{itemize}

The GDP is distinct from the Gross National Income (GNI),
which considers the nationality of the owners of productive
enterprises. In contrast, the GDP solely refers to the location
of production.

Both the GDP and the GNI are purely economic indices.
They do not account {\it e.g.}\ for the wealth distribution 
among the residents, the quality of health-care and education,
or the environment-friendliness of the production. The following two 
subsections will refer to some of these complementary aspects, 
which are more directly linked to the quality of human life.

\begin{figure}[h!]
\begin{center}
\hspace*{-6mm}
\includegraphics[width=0.355\textwidth,angle=270]{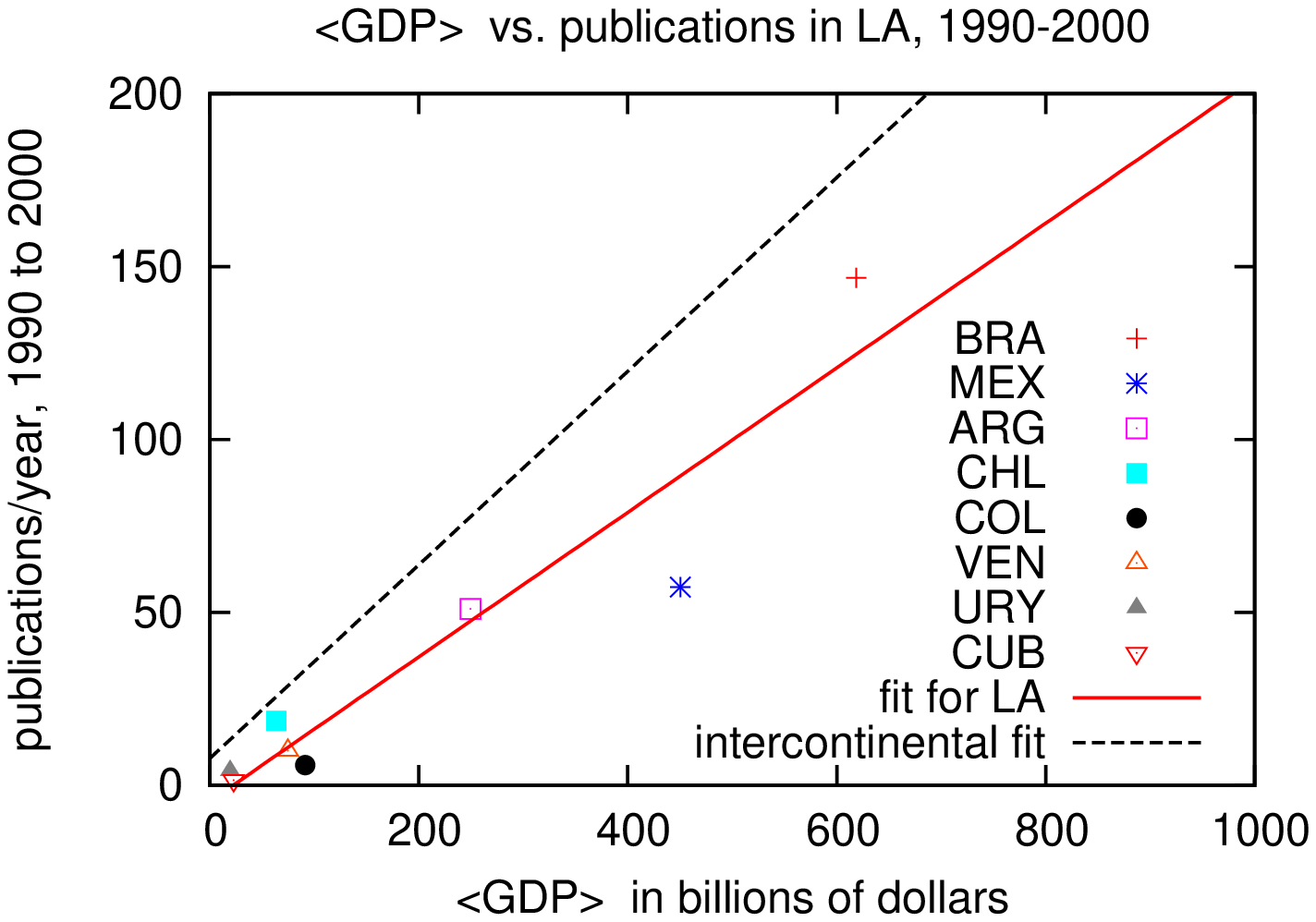}
\hspace*{-6mm}
\includegraphics[width=0.355\textwidth,angle=270]{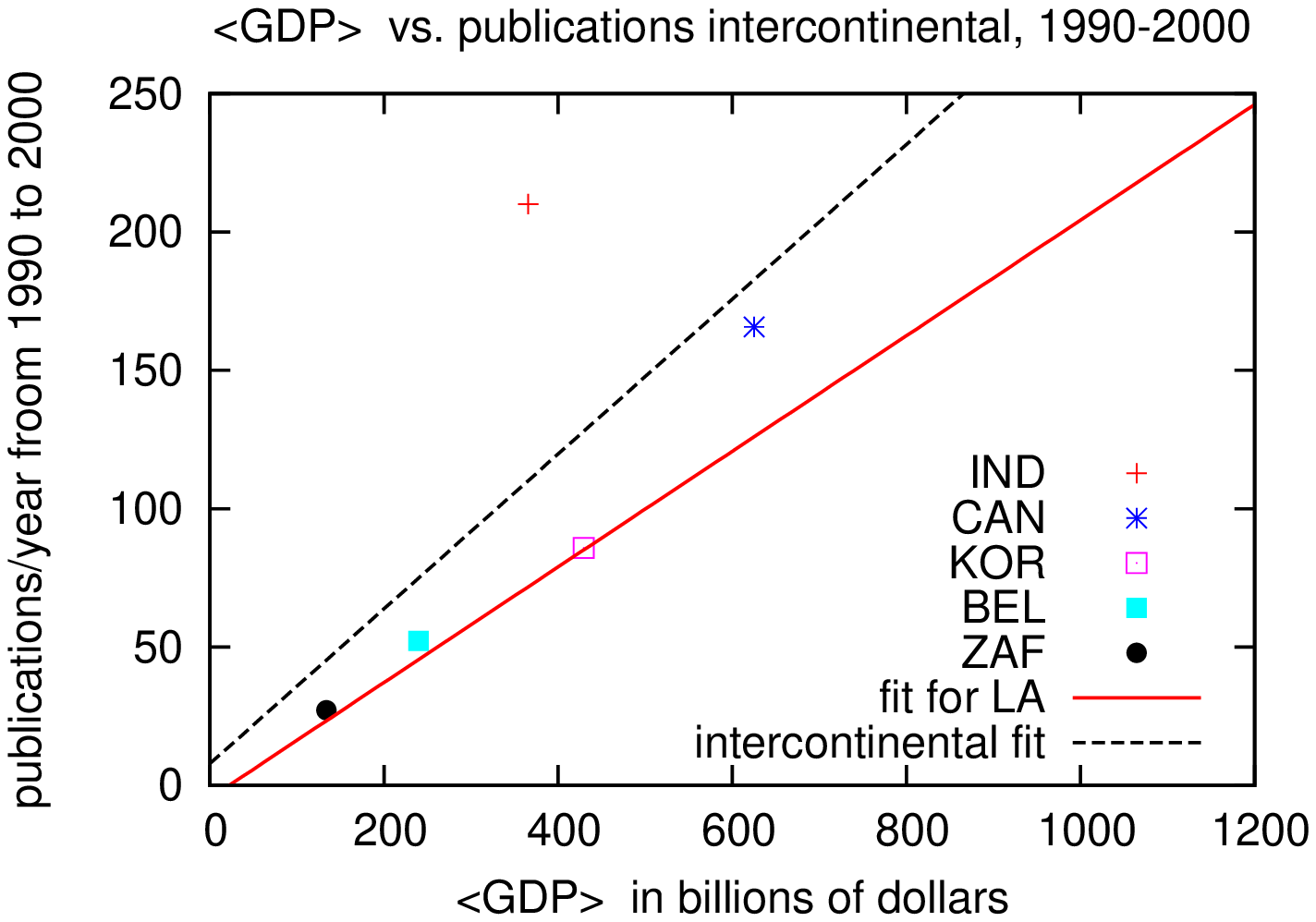} \\
\includegraphics[width=0.355\textwidth,angle=270]{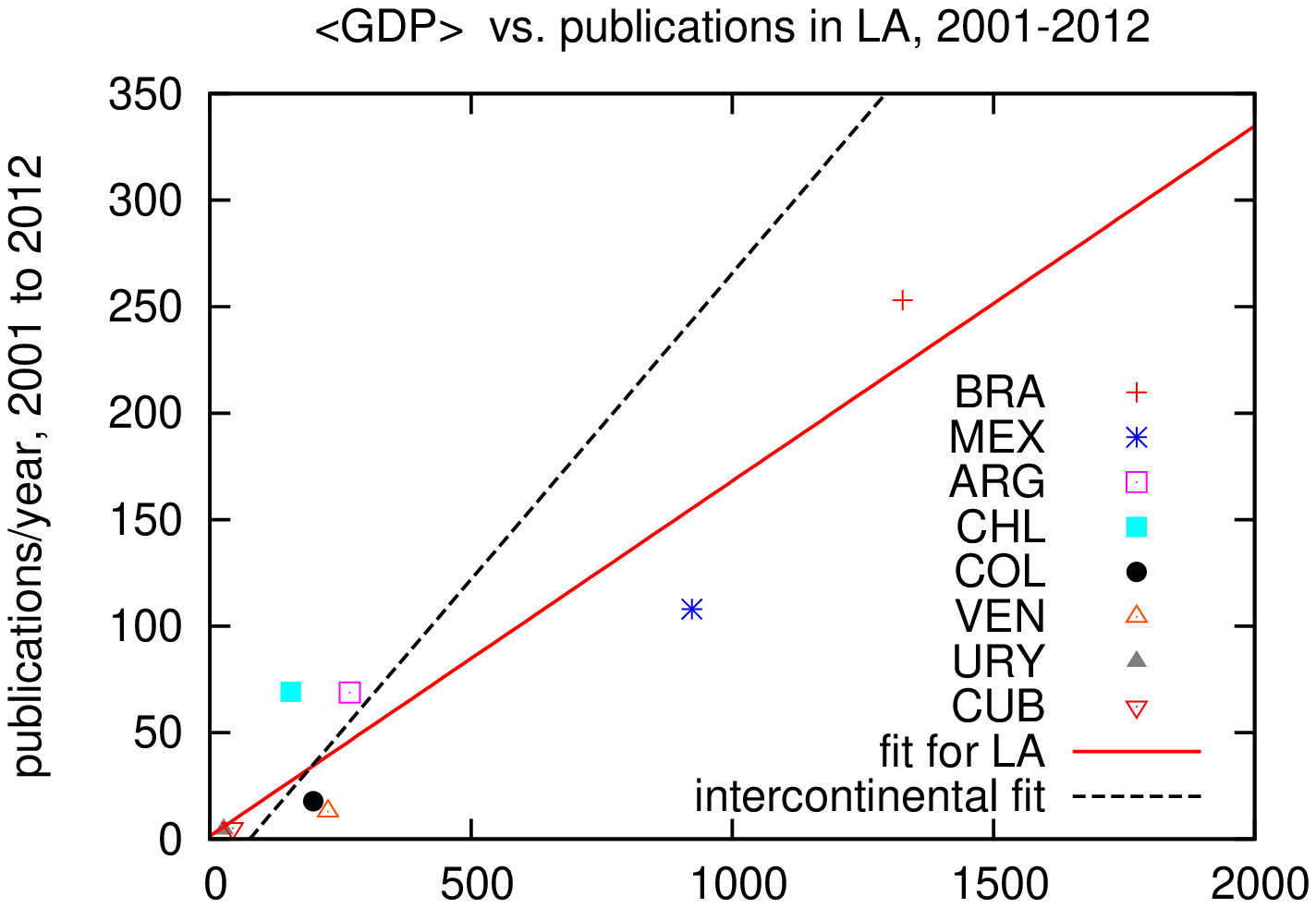}
\hspace*{-6mm}
\includegraphics[width=0.355\textwidth,angle=270]{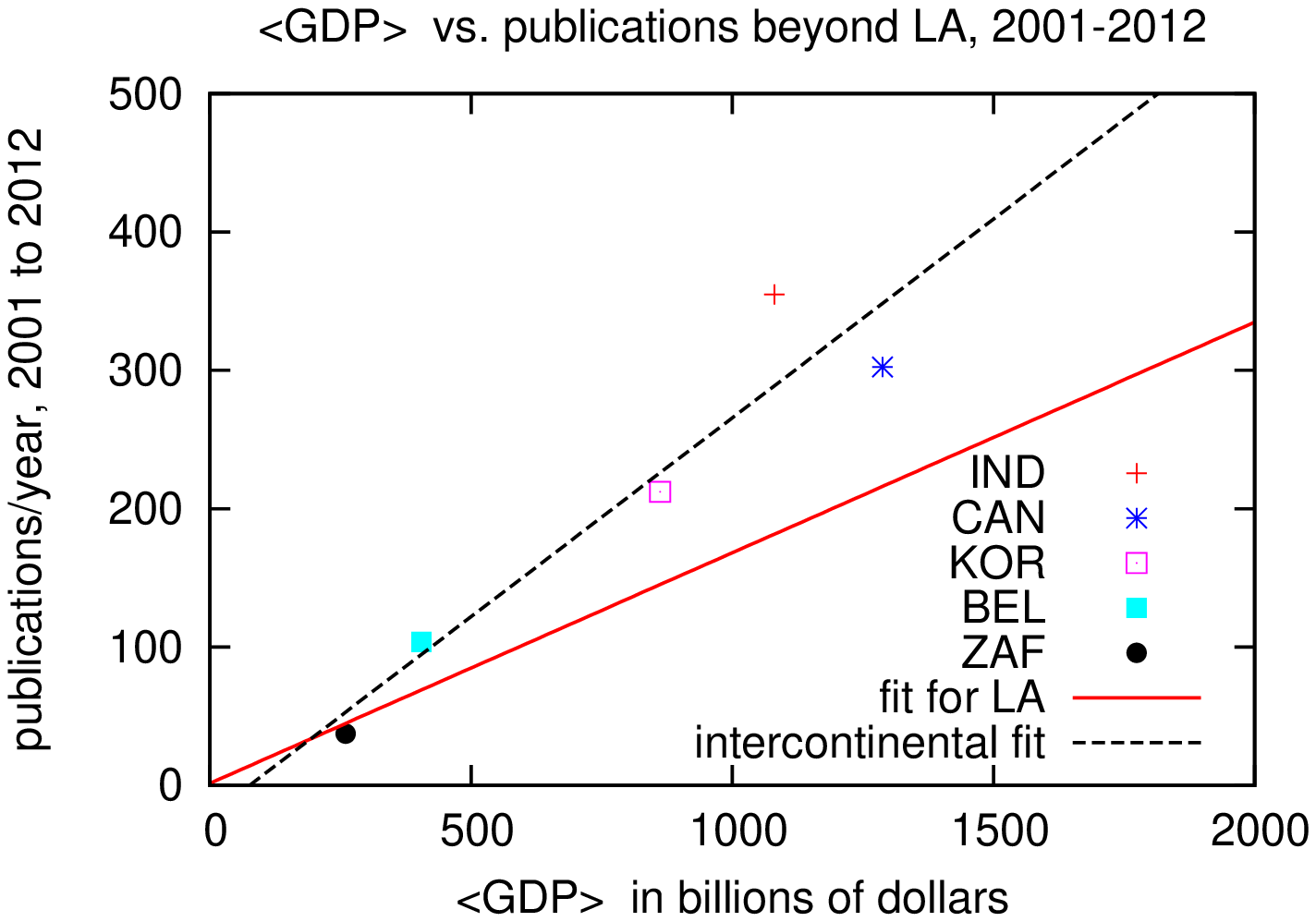}
\caption{The number of publication/year vs.\ the average GDP, in the
periods 1990-2000 (plots above) and 2001-2012 (plots below).
We show data for the countries under consideration
in LA (on the left) and beyond LA (on the right). For each of the two
periods we include linear fits for LA and for the intercontinental
countries. In both periods, the latter have a stronger slope: 
0.21 vs.\ 0.28 in the first period, and 0.17 vs.\ 0.29 in the second
period. This effect is mostly driven by India.
Within LA we see that particularly Brazil and Chile have a high
productivity of theoretical HEP publications relative to their GDP,
whereas the opposite holds for Mexico.}
\vspace*{-3mm}
\label{GDPpub}
\end{center}
\end{figure}

A study of the entire region of Latin America reports an increase
of the GDP by $50.8\,\%$ in the period 2002 to 2007, along with an
increase by $56.5\,\%$ of the Gross Domestic Expenditure for
Research and Development (GERD) \citep{PHealth}.\footnote{We do
  not have corresponding data specifically for HEP, let alone for
  theoretical HEP.}
This corresponds to an increase of the GDP world fraction from
$8.0\,\%$ to $8.5\,\%$; the GERD world fraction had a stronger
relative growth, though at a lower level, from $2.8\,\%$ to $3.5\,\%$.
\cite{impact} compare the world fraction of the GERD in
a number of countries to their impact, measured by the citation share
of papers in science and technology (averaged over 5 years).
In 1993 Chile had a balanced record (its GERD share and citation share
where both $0.1\, \%$), whereas in ARG, BRA, IND, MEX, ZAF the
citation share stayed behind the GERD share. This was still the
case in 2009, except for ARG, which now attained a balanced
record ($0.3\, \%$ each), while in Chile the citation share became
even stronger ($0.2\, \%$ citation share with still just
$0.1\, \%$ GERD share).

For earlier studies of the correlation between the GDP and
the overall scientific output in LA we refer to \cite{CarVil}
and \cite{LFJK} (for the 1980s) and \cite{MoyaHerrero}
(for the period 1991-7).
The latter reports an annual increase of the scientific production
by $13 \, \%$, with a ranking very similar to our Table \ref{pubtab}:
BRA, MEX, ARG, CHL, VEN, COL, CUB, URU. In addition to the correlation
with the GDP it considers the relation to the GERD,
and to the number of researchers in each country.

Figure \ref{GDPpub} refers to the two periods from 1990-2000,
and from 2001-2012. In both cases, we show for each country
in our study the total number of publications versus the
average GDP in this period. The data points do not follow
any obvious function, hence the only fit that appears sensible
is linear (although for instance India deviates
strongly form these fits).\footnote{Explicit data about the fit
  quality of the plots in this section are given in \cite{tesis}.}
It has been performed separately in and beyond LA, for both periods.
We observe a stronger slope for the intercontinental 
countries, {\it i.e.}\ if we extrapolate the publications/GDP from 
LA to the region of the intercontinental data points, the 
extrapolated line tends to be lower (South Africa is the exception). 
This trend is enhanced in the later period.

\subsection{Human Development Index (HDI) and Education Index (EI)}

The HDI was developed since the 1990s; key protagonists were
Mahbub ul Haq (from Pakistan) and Amartya Sen (from India).
Obviously, notions like well-being and happiness are hard to 
parameterize and to measure statistically.
However, a single index, which is more oriented towards the
quality of human life than the GDP, is motivated by the purpose
of shifting the attention of influential persons (such as 
politicians) from purely economic criteria towards a more 
social perspective.

\begin{figure}[h!]
\begin{center}
\hspace*{-6mm}
\includegraphics[width=0.355\textwidth,angle=270]{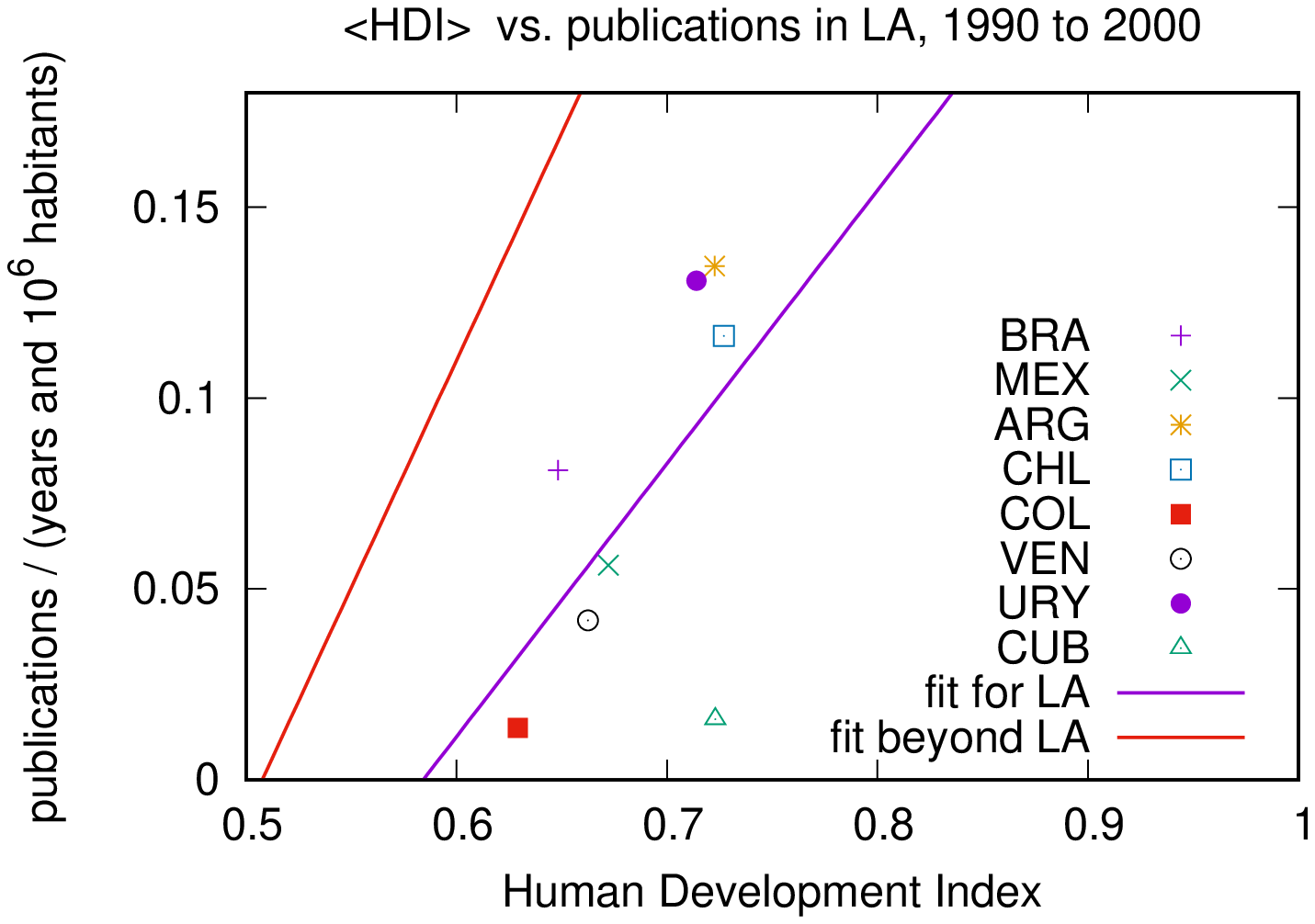}
\hspace*{-6mm}
\includegraphics[width=0.355\textwidth,angle=270]{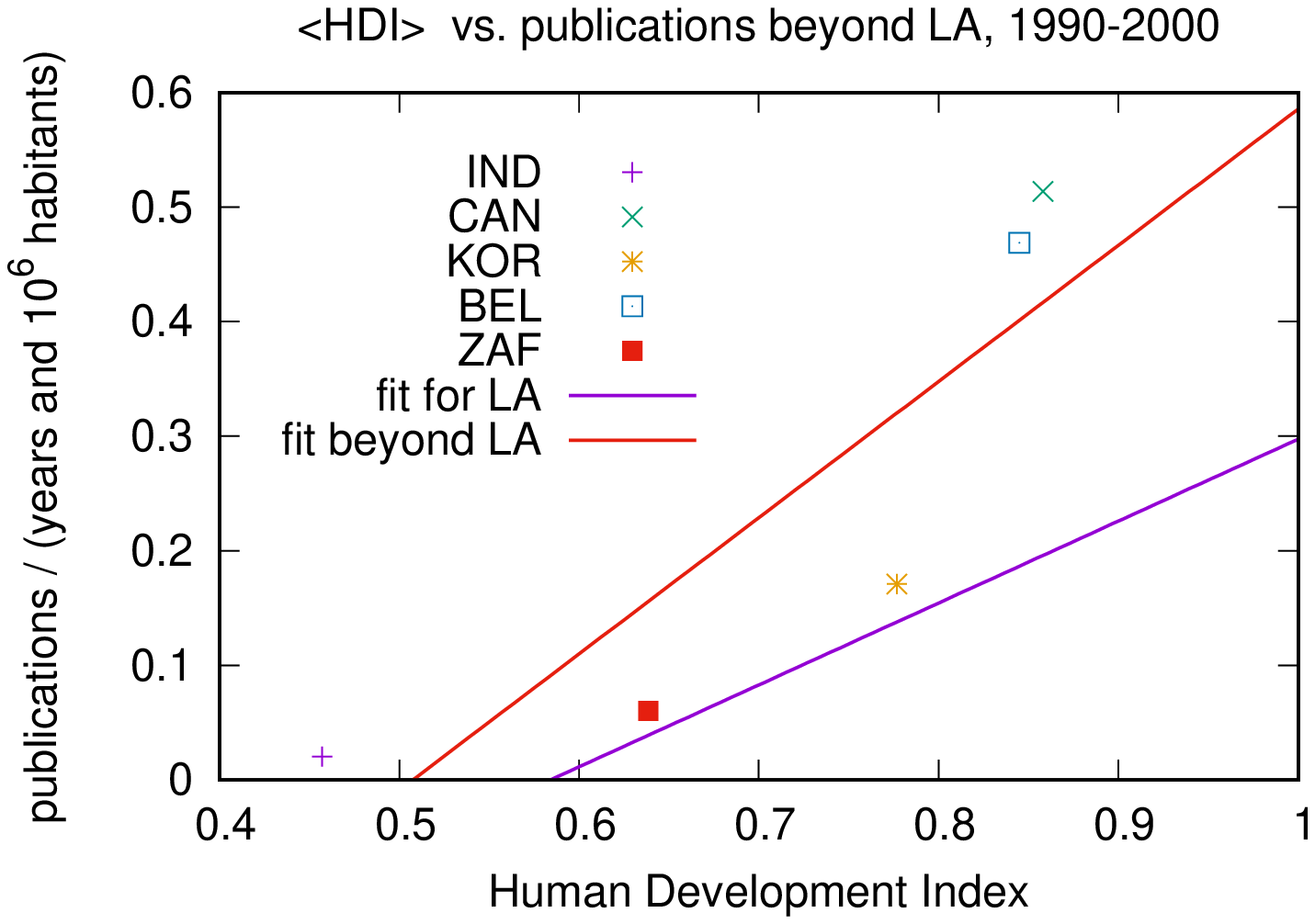} \\
\includegraphics[width=0.355\textwidth,angle=270]{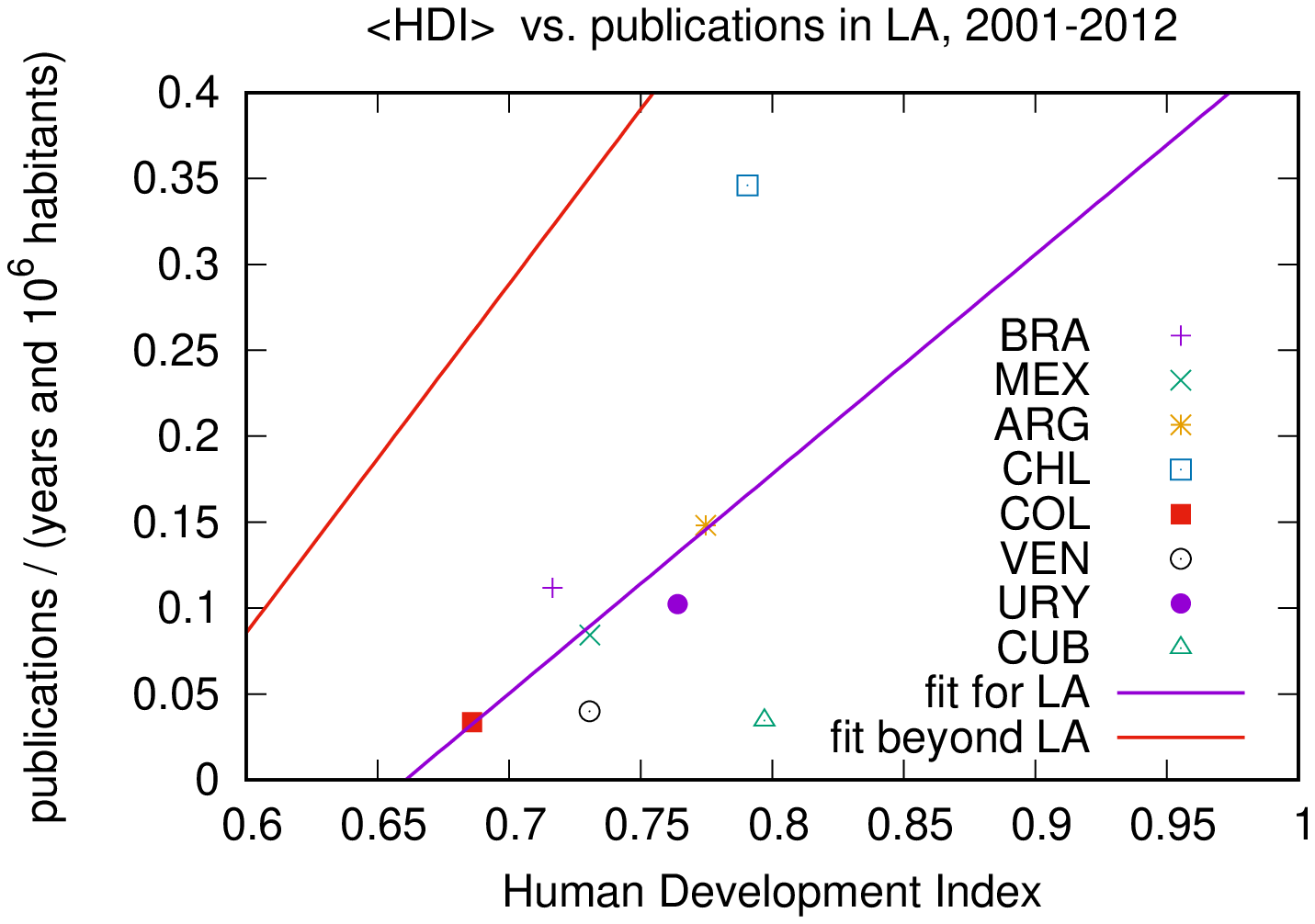}
\hspace*{-6mm}
\includegraphics[width=0.355\textwidth,angle=270]{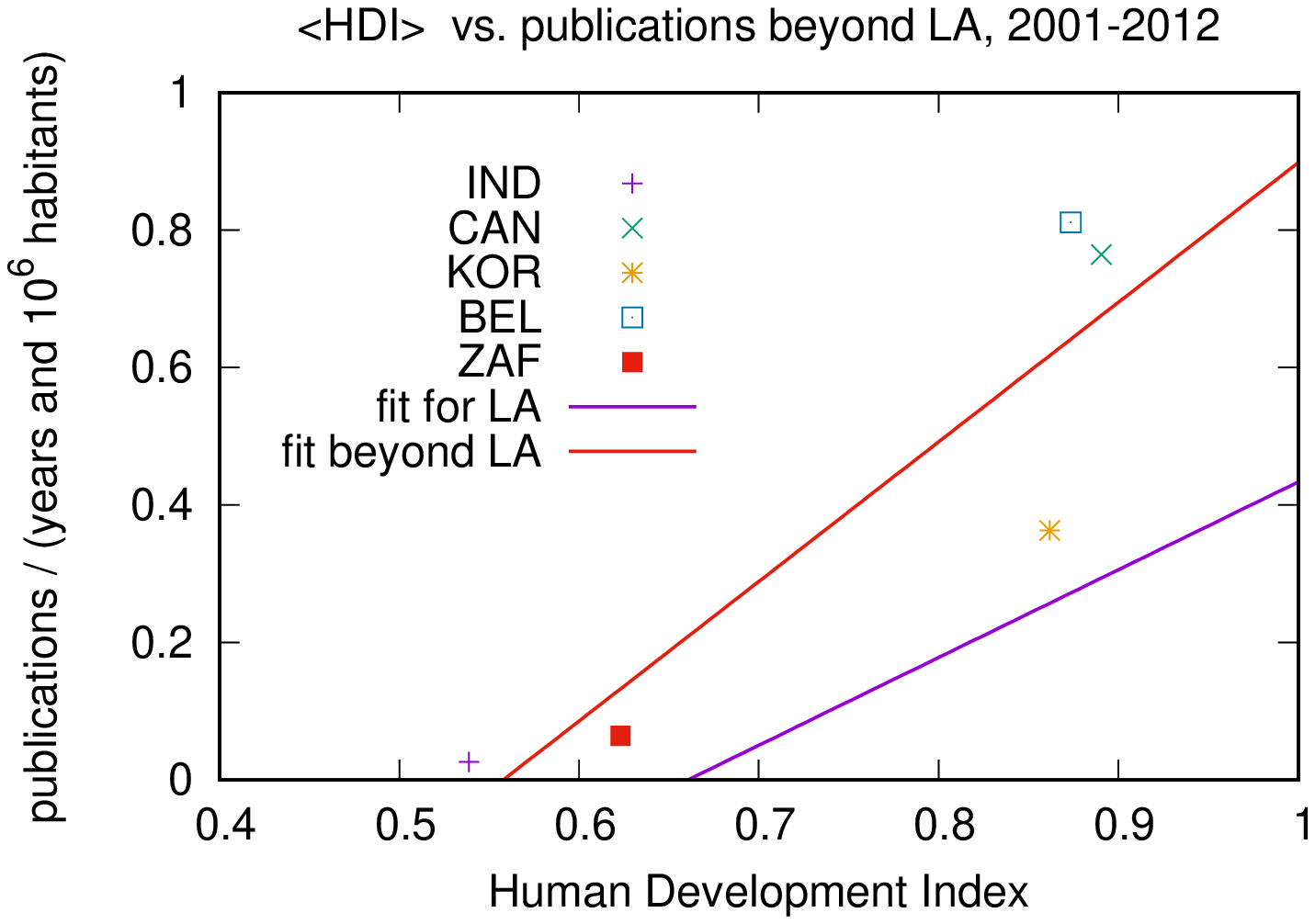}
\caption{The number of publications/(million of inhabitants $\times$ year)
vs.\ the average HDI, in the period 1990-2000 (plots above)
and for the period 2001-2012 (plots below).
We show data for the countries under consideration
in LA (on the left) and beyond (on the right). For each of the two
periods we include linear fits for the LA and for the intercontinental
countries --- although these fits have a rather poor quality.
In both periods, the latter have a stronger slope: 
0.72 vs.\ 1.19 in the first period, and 1.28 vs.\ 2.03 in the second
period. This effect is due to Canada, Belgium and India, while the
publication/HDI ratio for South Korea and South Africa is similar to 
LA. Within LA, Brazil and Chile have a high HEP productivity
relative to their HDI, while the opposite holds for Cuba.}
\vspace*{-3mm}
\label{HDIpub}
\end{center}
\end{figure}

The HDI is composed of three indices, which are related to
health, education and income \citep{HDI}. Before 2010 these three 
components were computed as follows:

\begin{itemize}

\item Life Expectancy Index LEI = $\frac{{\rm LE}-25}{60}$,
where LE is the life expectancy (in years).

\item Education Index EI = $\frac{2 \times {\rm ALI} + {\rm GEI}}{3}$,
  where Adult Literacy Index ALI =
adult literacy and Gross Enrollment Index GEI =
school enrollment (both in fractions of $1$).

\item Income Index II = $\frac{\ln ({\rm GDPpc}/100)}{\ln 400}$,
where GDPpc = GDP per capita.

\end{itemize}

This assumes an effectively maximal LEI and II of 1 for 85 years
of life expectancy, and a GDPpc of 40$\,$000 dollars.
Finally the HDI is obtained as the geometric mean of these three
indices:
$
{\rm HDI} = ({\rm LEI} \times {\rm EI} \times {\rm II} )^{1/3} \ . 
$
Compared to the arithmetic mean, it decreases the more the three
components are distinct.

In 2010 the definitions of the EI and the II (but not of the LEI)
were refined:

\begin{itemize}

\item EI = $\frac{1}{2} ({\rm MYSI} + {\rm EYSI})$, 
where MYSI = 
(mean years of schooling of the populations above 24 years)$/15$,
and EYSI = (expected years of schooling of a 5-years old child)$/18$.
(The denominator 18 corresponds to the usual duration of the education
up to a master's degree.)

\item II = $\frac{\ln ({\rm GNIpc}/100)}{\ln 750}$
  where GNIpc = Gross National Income per capita,
  at purchasing power parity (cf.\ Subsection 4.1).

\end{itemize}

\begin{figure}[h!]
\begin{center}
\hspace*{-6mm}
\includegraphics[width=0.355\textwidth,angle=270]{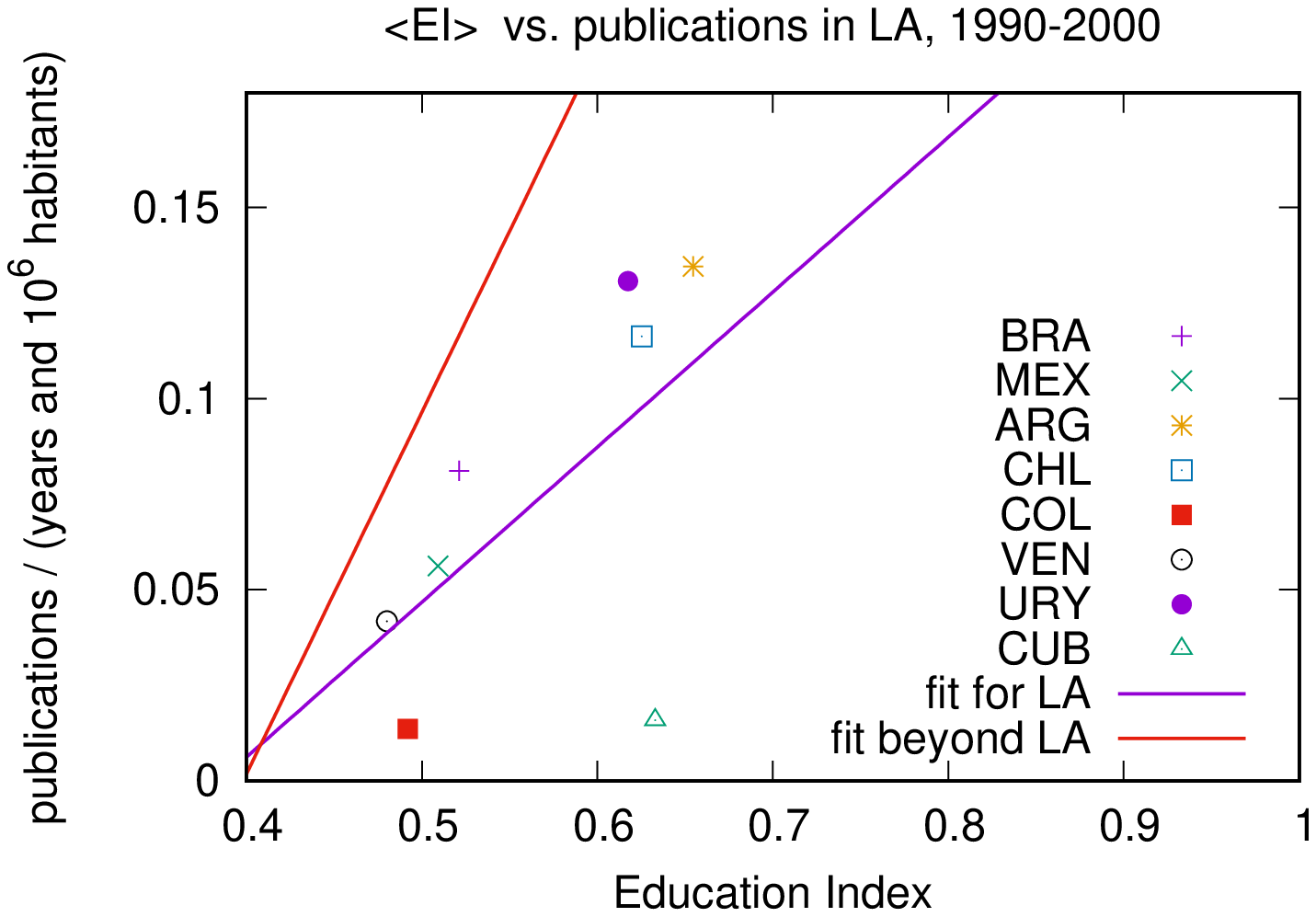}
\hspace*{-6mm}
\includegraphics[width=0.355\textwidth,angle=270]{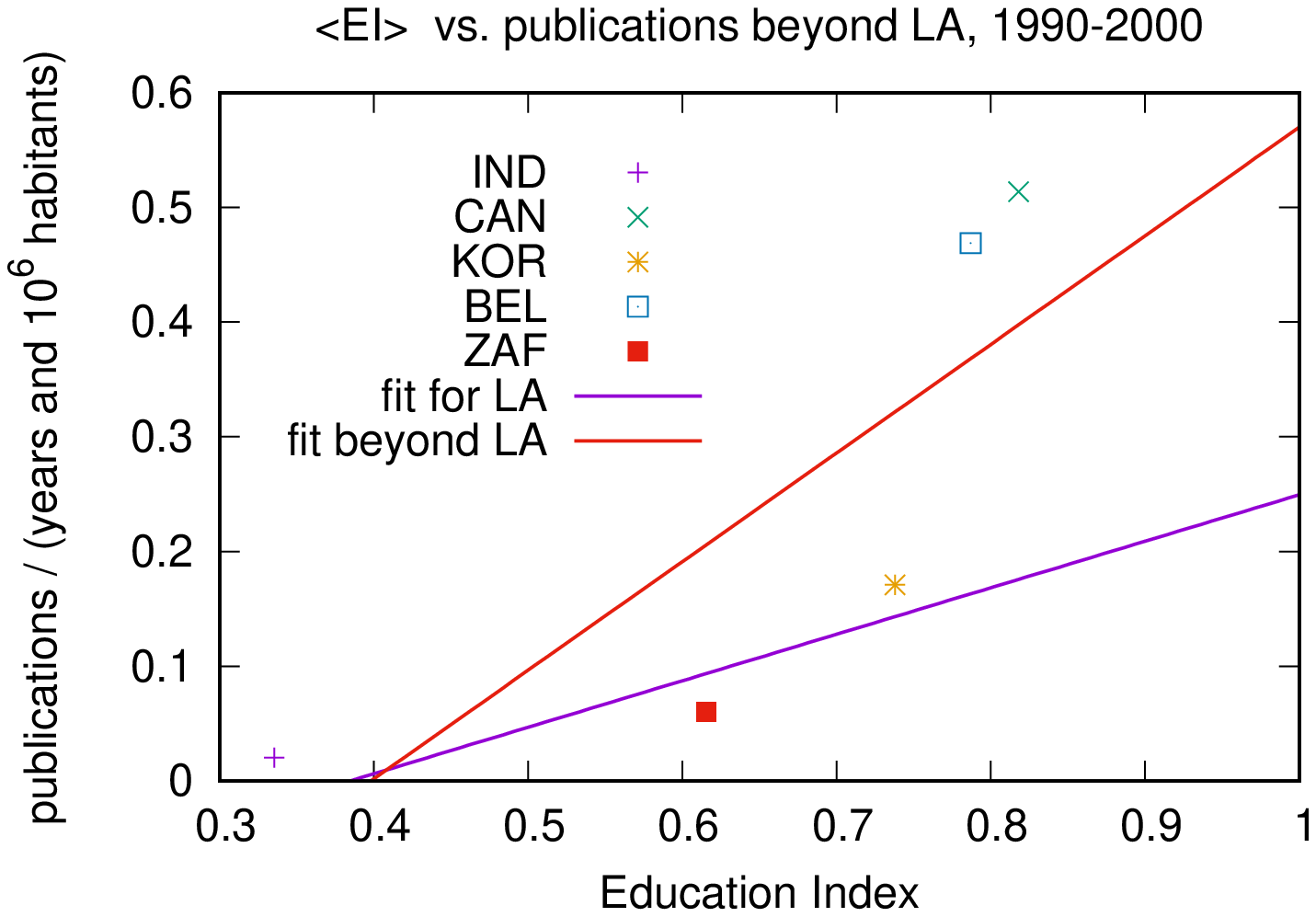} \\
\includegraphics[width=0.35\textwidth,angle=270]{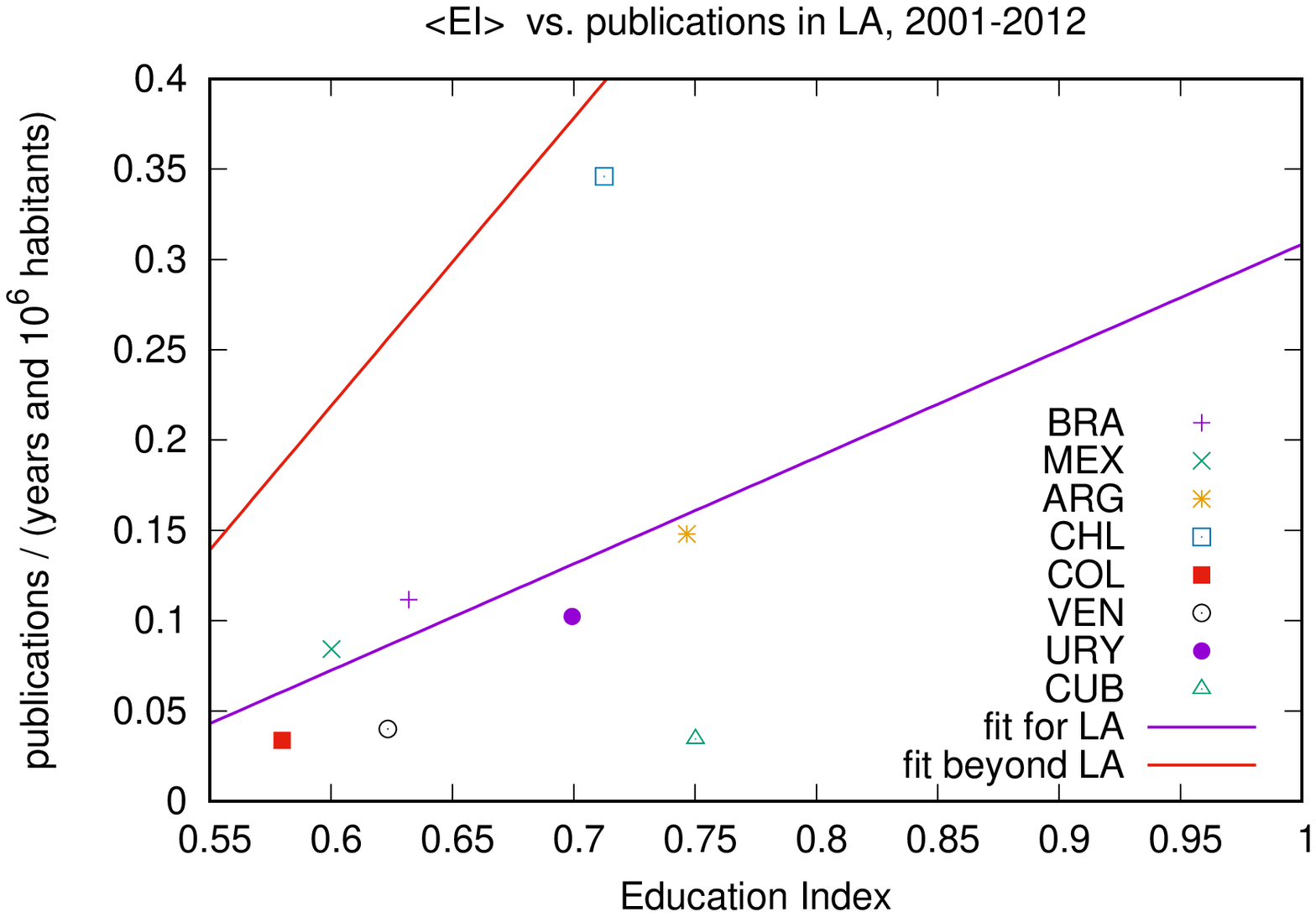}
\hspace*{-6mm}
\includegraphics[width=0.35\textwidth,angle=270]{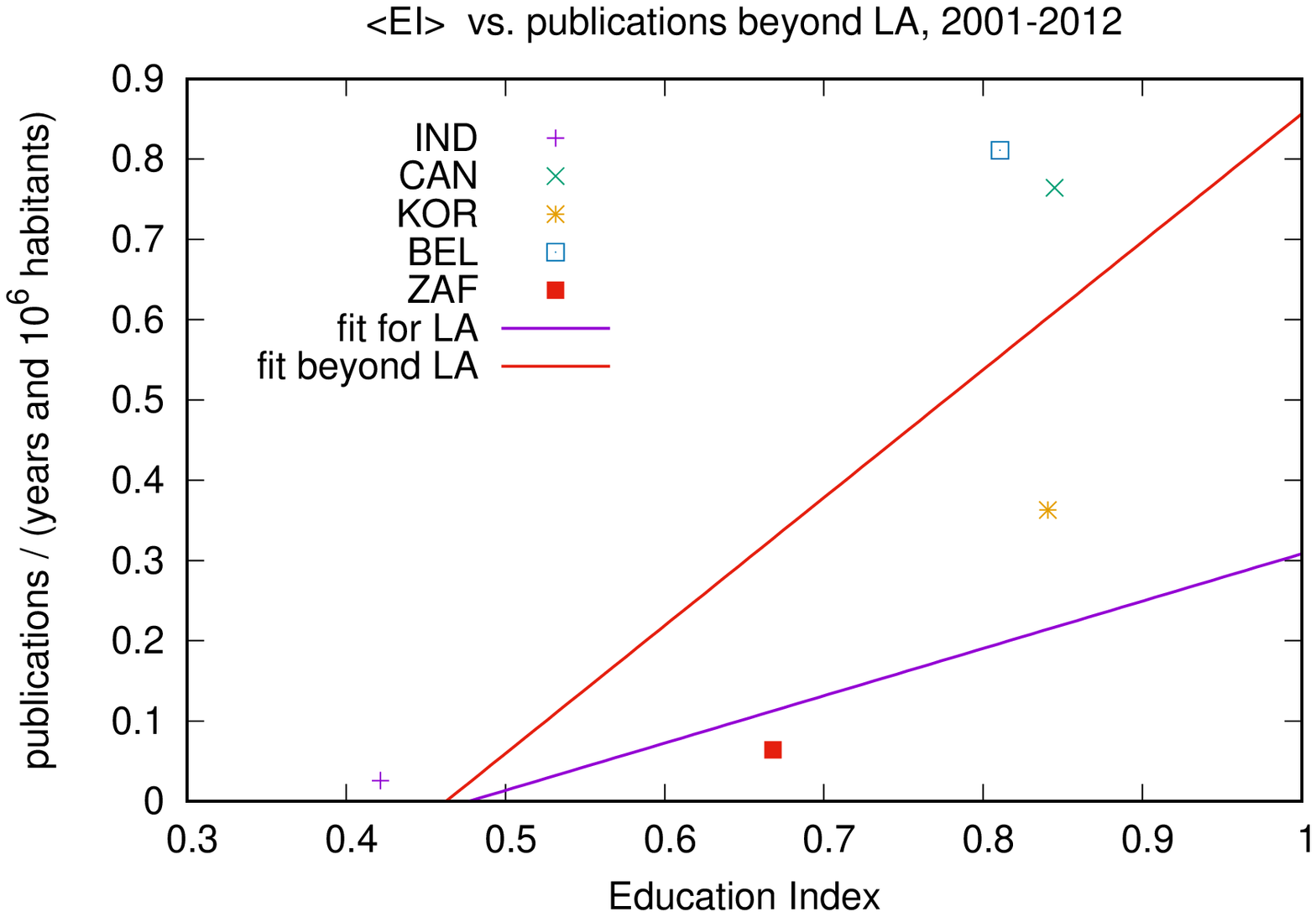}
\caption{The number of publications/(million of inhabitants $\times$ year)
vs.\ the average EI, in the period 1990-2000 (plots above)
and for the period 2001-2012 (plots below).
We show data for the countries under consideration
in LA (on the left) and beyond (on the right). For each of the two
periods we include linear fits for the LA and for the intercontinental
countries. In both periods, the latter have a stronger slope: 
0.41 vs.\ 0.95 in the first period, and 0.59 vs.\ 1.59 in the second
period. The observations are very similar to Figure \ref{HDIpub},
which refers to the HDI (the EI is one of its components).}
\vspace*{-3mm}
\label{EIpub}
\end{center}
\end{figure}

Figure \ref{HDIpub} compares the HDI to the number of
publications per million of inhabitants and year.
In addition, Figure \ref{EIpub} refer exclusively to the EI,
which can be viewed as the direct basis for scientific research.

As in Figure \ref{GDPpub}, the plots in Figures \ref{HDIpub} and
\ref{EIpub} are divided into two periods, and without any obvious
fitting function, a linear interpolation seems to be again the
only applicable fit. In both periods, and in both figures,
the fits for the intercontinental countries have a stronger
slope. This means that in those countries the given conditions
--- the HDI and EI --- are converted into a higher production of
physics papers that enter our statistics. The same qualitative
feature was observed for the GDP in Figure \ref{GDPpub}.

\subsection{Search for correlations in the time-evolution}

We proceed to a schematic overview of the time-line of the
socio-economic indices, searching for trends, where a relation 
to the productivity of theoretical HEP papers is conceivable.
Here we refer to properties, which would be called ``intensive''
in the terminology of thermodynamics, {\it i.e.} they do not
depend on the size, or in our case the total population of
a country. In particular, we consider the
publication rate {\em per million of inhabitants,} which
modifies the leading and sub-leading group, as Table
\ref{pubpoptab} shows. Correspondingly, we now consider the
GDP {\em per capita} (GDPpc), as well as HDI and EI, which
are ``intensive'' as well. (In this terminology, the total
number of publications or citations and the GDP would be
denoted as ``extensive'' quantities).

\begin{figure}[h!]
\begin{center}
\includegraphics[width=0.48\textwidth,angle=0]{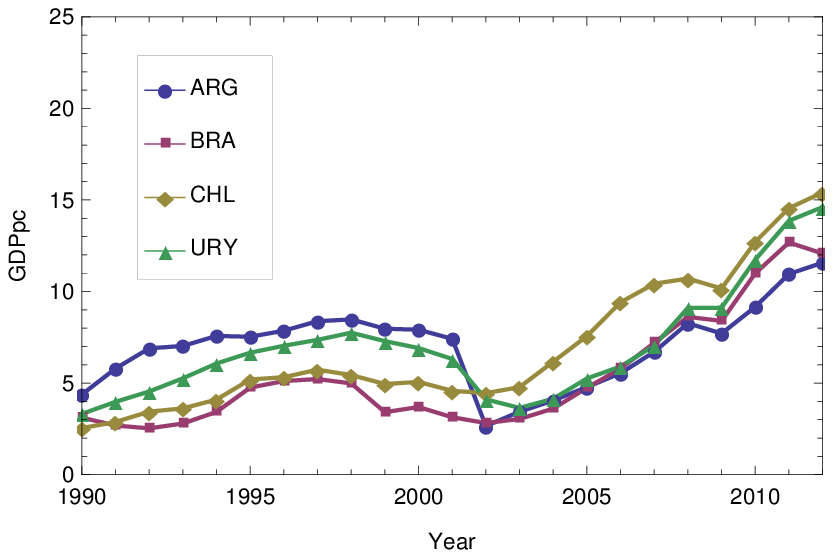}
\includegraphics[width=0.48\textwidth,angle=0]{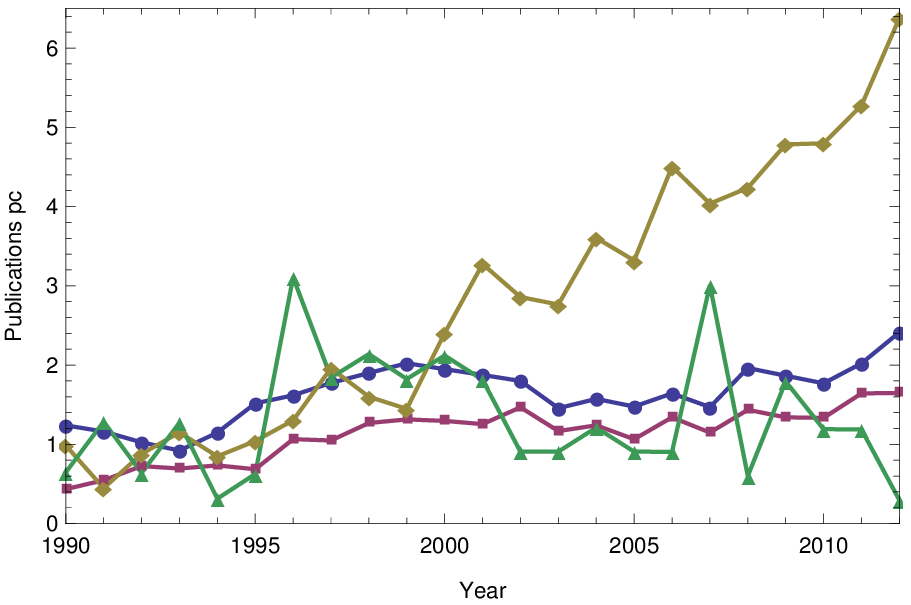} \\
\includegraphics[width=0.49\textwidth,angle=0]{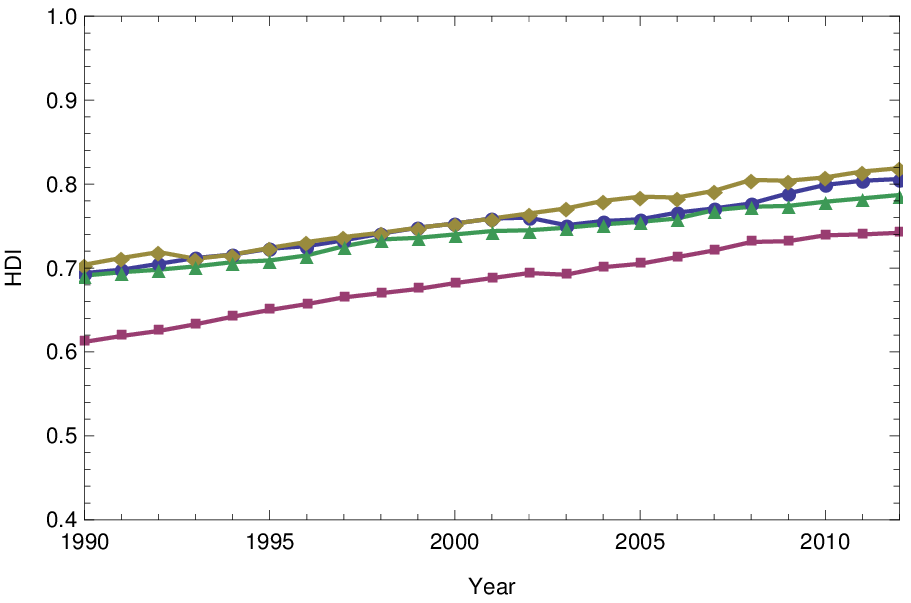}
\includegraphics[width=0.49\textwidth,angle=0]{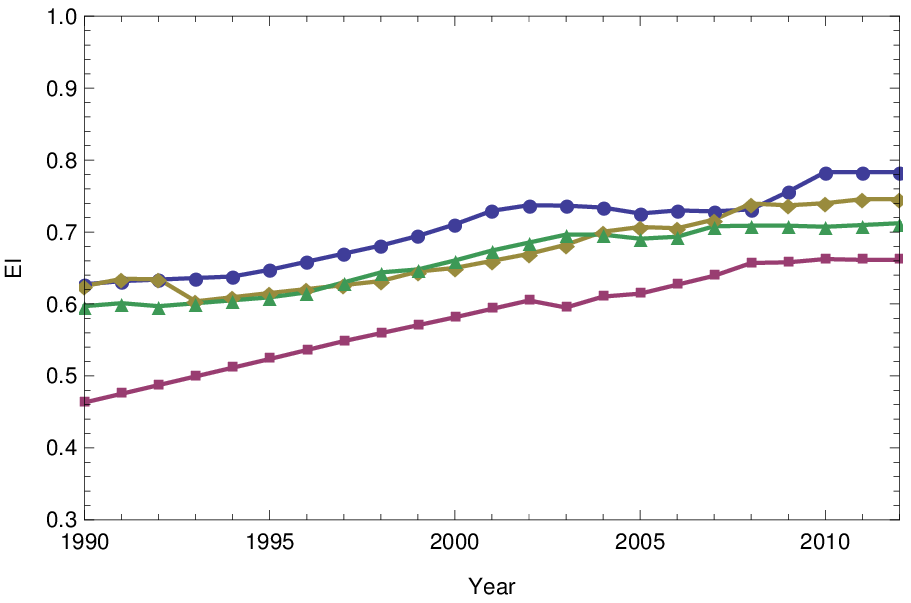}
\caption{The time-line of the GDPpc (per capita, in units of 1000
  dollars), the publication rate per million of inhabitants, the
  HDI and the EI of the leading group of LA countries --- with respect
  to the theoretical HEP productivity per capita --- in the period 1990-2012.}
\vspace*{-5mm}
\label{timelineLat1}
\end{center}
\end{figure}

\begin{figure}[h!]
\begin{center}
\includegraphics[width=0.48\textwidth,angle=0]{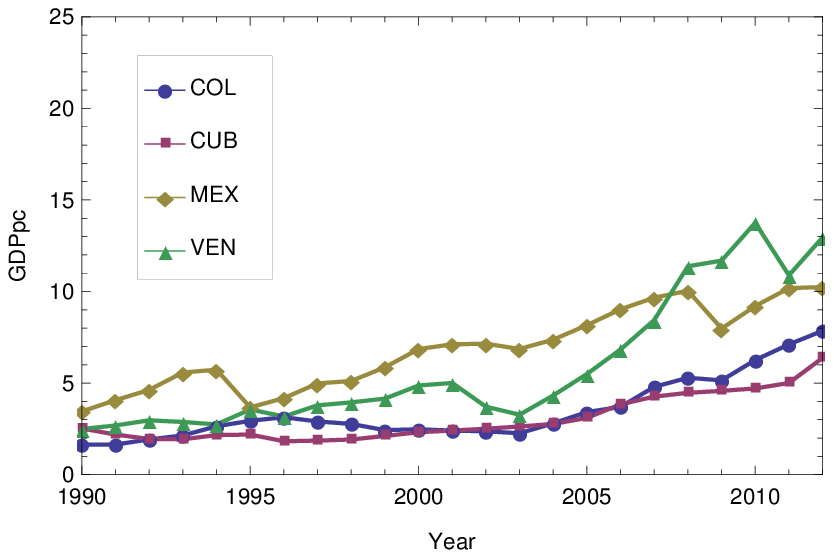}
\includegraphics[width=0.48\textwidth,angle=0]{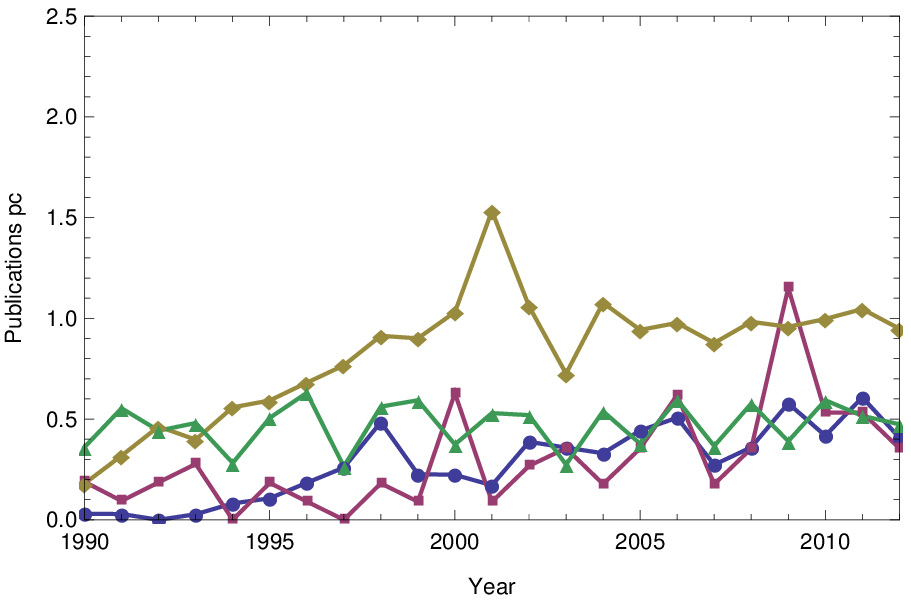} \\
\includegraphics[width=0.49\textwidth,angle=0]{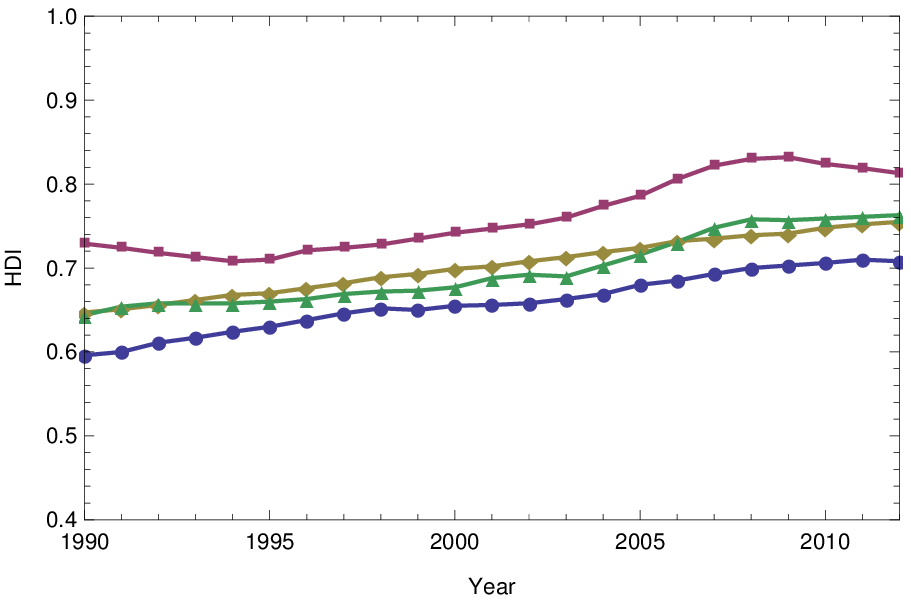}
\includegraphics[width=0.49\textwidth,angle=0]{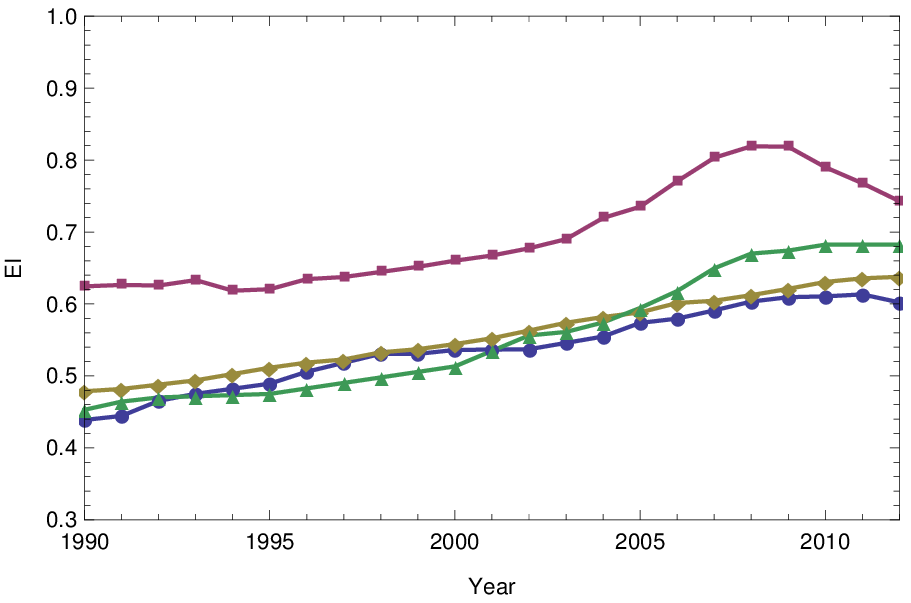}
\caption{The time-line of the GDPpc (per capita, in units of 1000 dollars),
  the publication rate per million of inhabitants, the HDI and the
  EI of the sub-leading group of LA countries --- with respect to
  the theoretical HEP productivity per capita --- in the period 1990-2012.
  (Note that the scale for publication pc differs from
  Figures \ref{timelineLat1} and \ref{timelineInt}.)}
\vspace*{-5mm}
\label{timelineLat2}
\end{center}
\end{figure}

\begin{figure}[h!]
\begin{center}
\includegraphics[width=0.48\textwidth,angle=0]{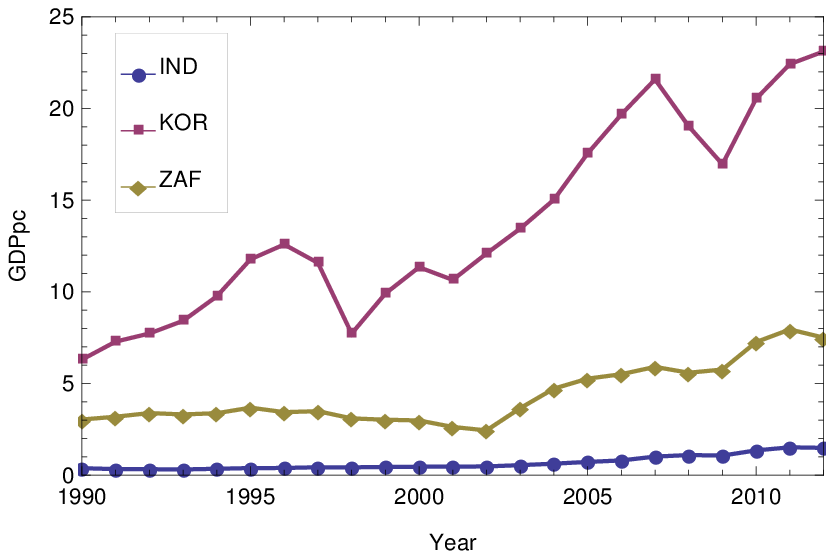}
\includegraphics[width=0.48\textwidth,angle=0]{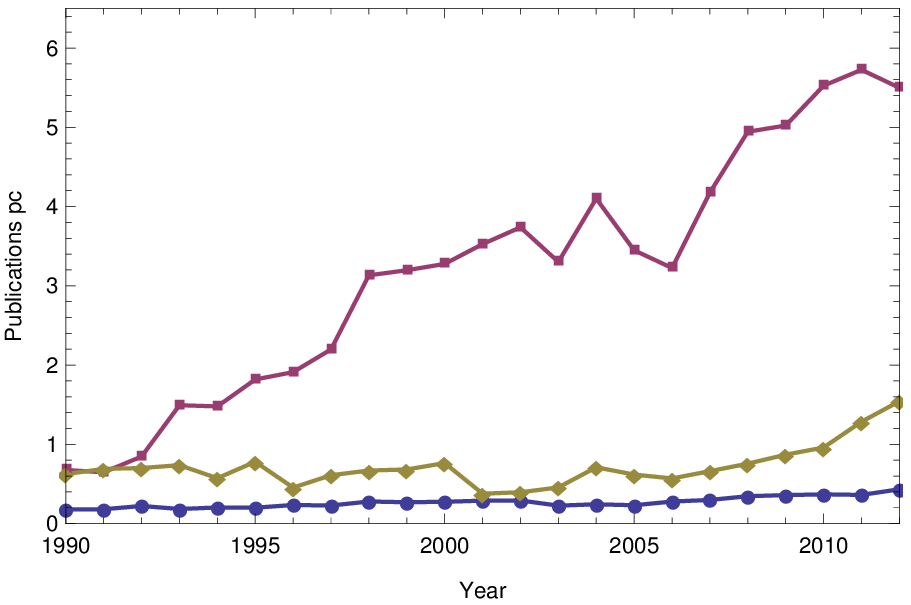} \\
\includegraphics[width=0.49\textwidth,angle=0]{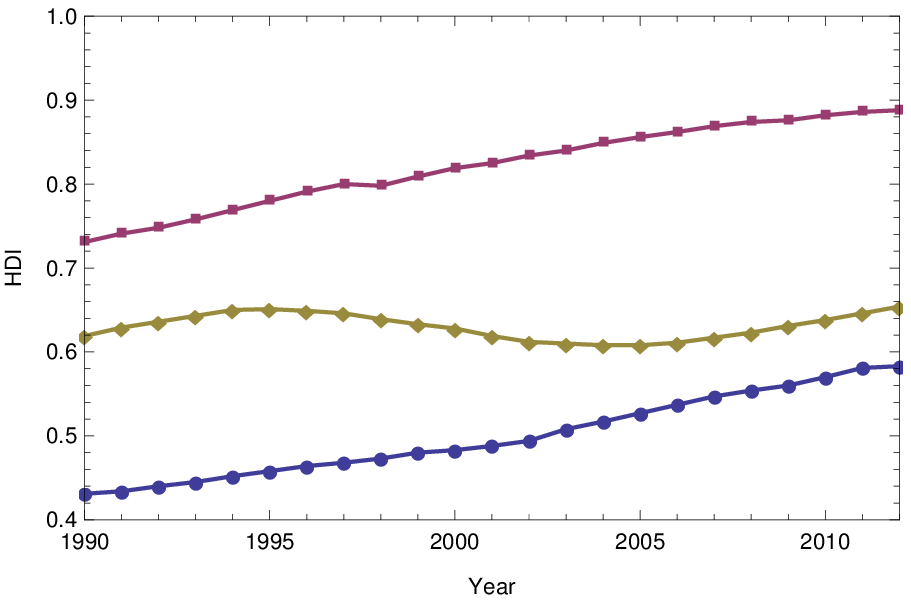}
\includegraphics[width=0.49\textwidth,angle=0]{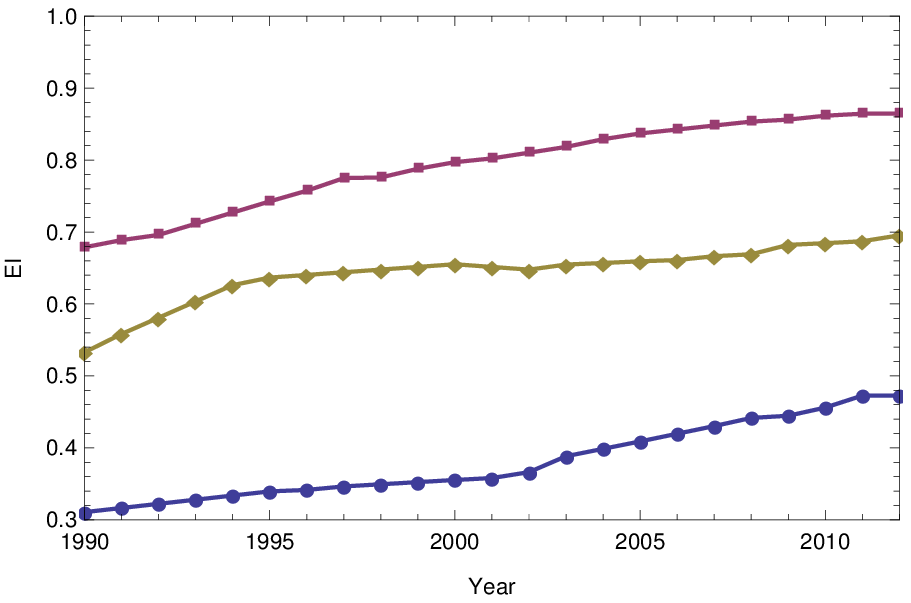}
\caption{The time-line of the GDPpc (per capita, in units of 1000 dollars),
  the publication rate per million of inhabitants, the HDI and the
  EI in three selected countries outside LA, in the period 1990-2012.}
\vspace*{-5mm}
\label{timelineInt}
\end{center}
\end{figure}

The time-lines of these four intensive quantities are shown
in Figures \ref{timelineLat1}, \ref{timelineLat2} and \ref{timelineInt}.
In several countries we observe a marked {\em economic recession} in the
period 2002 to 2003, but afterwards a {\em significant growth.} This
feature is extreme in {\em Argentina,} where in 2002 the GDPpc collapsed
down to $36\, \%$ of its value in 2001. It seems obvious to relate this
crises to the subsequent period of stagnation (including slight
recessions) of the HDI, EI and publications per capita, which lasted
for about 5 years.

An analogous, but less extreme, economic crises around 2002,
followed by significant growth, is observed in Uruguay and
Venezuela, while in Brazil, Chile, Colombia, South Africa and
South Korea the growth after 2003 coincides.
In {\em Uruguay} a stagnation of the EI after 2002
could be a consequence of the economic crises, but no detailed
correlation with the publication number is visible.\footnote{Here
  the annual publication number per capita fluctuates strongly.
  This can be explained by the relatively small population, so
  the HEP publications are written by very few individuals. For the
  same reason, the publication rate fluctuates strongly also in Cuba
  and Chile.}
{\em Venezuela} achieved a relevant improvement in its HDI and
EI in the early 21st century, but the publication number did not
follow this trend --- on a long-term scale it is stable.
{\em Colombia} improved its HDI and EI in this period
(but it is still at a modest level), while its publication
rate was slightly increased, but there are no correlated fluctuations.

We mentioned already that the economies of Brazil and Chile
strongly expanded after 2003. In this period, the
HDI and EI increased as well, but again there is no detailed
correlation. {\em Chile} attained the strongest increase in
publications per capita, so it clearly dominates in Latin America
in this respect, although its EI is below Argentina and Cuba.
In {\em Brazil} the publication rate per capita increased
marginally, without really reflecting the strong boost in the GDPpc.

The {\em Mexican} economy followed a more continuous growth (except for
the years 1995 and 2009), the same holds for the HDI, EI, and also
the publication rate had a weak but mostly continuous trend up (except
for fluctuations in 2001 and 2003). So all four indicators followed
a similar behavior, but there is no evidence for a detailed
correlation. The {\em Cuban} economy grew gradually in the 21st
century, while the HDI and EI were strongly boosted (they decreased after
2009, but still at a high level). The publication number fluctuates
strongly, with a long-term trend up. It would be a matter of speculation
to relate the peak in 2009 to the maxima of the HDI and EI.

Beyond Latin America, {\em South Korea} exhibits a strong (though not always
continuous) economic growth, which seems consistent with an improvement
of the other three parameters under consideration. The evolution
in {\em India} is trend-wise similar, but at a much lower level.
In {\em South Africa} the GDPpc clearly increased after 2003,
as in a number of Latin American countries.
The publication number increased as well in this period, although
the HDI does not follow this trend. 

To summarize, a long-term trend up in all four parameters is manifest
everywhere. Beyond this general pattern, a relation of the
socio-economic circumstances and the publication rate seems
obvious in the case of Argentina, and it might be plausible
in Chile and South Korea, but in most cases such a detailed
correlation cannot be observed.

\section{Conclusions}

We presented a statistical analysis of the activity in 
theoretical HEP in Latin America, in the period from 1990
to 2012. We considered the 8 dominant countries in LA, as
well as 5 countries in other regions for comparison.

Regarding the number of publications and citations,
the leading nations in LA are Brazil, Mexico, Argentina
and Chile. If we consider this productivity relative to 
the population, the hierarchy reads: Chile, Argentina,
Uruguay, Brazil.

As a generic trend, the productivity increases with time.
More wealth of a country --- measured in terms of its GDP
--- tends to increase its productivity in theoretical HEP, but 
the correlation does not follow any simple rule, nor detailed
fluctuations in time.
An obvious interpretation is that the activity in theoretical HEP 
significantly depends on the priorities in a country, even 
within science, and not only on its economic potential.\footnote{Our
ranking can be compared to a branch of science, which is devoted to
practical benefits (in contrast to HEP), like public health.
A study of publications in this field \citep{PHealth}
identified 10 leading countries in LA, which coincide
with the countries in our list, plus Peru and Puerto Rico.
In this field, Brazil strongly dominates with a production
$67.25\,\%$ of all LA publications, followed by MEX, COL,
CHL, CUB and ARG, {\it i.e.}\ particularly COL and CUB
are stronger than in HEP. From an intrinsic perspective,
relative to the population, the ranking changes more
drastically: here {\em Cuba} is the leader, followed by Puerto Rico,
BRA, CHL and URU. This illustrates differences in the national
research priorities, like the Cuban focus on medical research.}

Nevertheless a linear fit of the national data for the number of
publications vs.\ the GDP appears reasonable.
The comparison shows a stronger slope in the intercontinental 
countries, in particular in the 21st century. Therefore,
in order to explain the lower productivity in LA compared to
the leading countries in Asia, Europe and North America, it is
{\em not} sufficient to refer to the lower economic potential in LA.
If the interpolation of the intercontinental countries 
is linearly extended to the GDPs in LA, the extrapolated
productivity is still above the one observed in LA.

For instance, \cite{MoranLopez} refers to ``poor economic
conditions in most Latin American countries'' as ``one major reason''
why this subcontinent is not as productive in physics as the ``developed
countries in the North''. This is certainly correct, but our study
suggests that this reason alone is not a sufficient explanation. 

The observation of a superior (linear) extrapolation of the
HEP productivity outside LA persists similarly for
the HDI, and in particular for the EI (as one of its components).
Again a monotonous dependence can be seen, and a linear dependence 
is the best (reasonable) fit, but some countries deviate strongly
from this interpolation. 

As for the detailed time-evolution, there are in most cases no
simple relations between the publication rate and socio-economic
indices. An exception is the economic collapse of Argentina in 2002,
which did affect theoretical HEP. In a few other countries, including
Chile, a scientific boost due to an economic boom is conceivable.
However, such an impact is not a generic rule (our discussion includes
counter-examples), so in general the publication rate does not
react in an obvious manner to the socio-economic development.

\vspace*{-5mm}

\ \\
\ \\
{\bf Acknowledgements} \ 
We thank Magdalena Sierra for technical assistance, and the reviewers
for attracting our attention to a number of relevant references.
This work was supported by the Mexican Consejo Nacional de Ciencia
y Tecnolog\'{\i}a (CONACYT) through project CB-2010/155905, and by 
DGAPA-UNAM, through grant IN107915 and through the program PASPA-DGAPA.

\appendix

\section{Famous Latin American publications in theoretical high energy physics}

\subsection{Top-cited LA papers in theoretical HEP, published
  in specialized journals between 1990 and 2012,
  regarding citations until 2012}

We add a list of the eight most cited LA papers
in our statistics (each of them can be assigned to exactly one
LA country). Each of these eight papers has over 250 citations,
which means that it is {\em famous} according to the terminology
used in the online data base \cite{inspire}, which 
is freely accessible and highly popular in HEP.
However, our citation numbers added to this list are counted
in the Web of Science, up to the end of 2012, as described in
Section 2.\\

\newcommand{\up}{\vspace*{2mm}}

\noindent
[1] M.\ Banados, M.\ Henneaux, C.\ Teitelboim and J.\ Zanelli, 	
``Geometry of the (2+1) black hole'', Phys.\ Rev.\ D48 (1993) 1506
[{\underline{Chile}}] 799 citations.
\up

\noindent
[2] R.\ Gambini and J.\ Pullin,
``Nonstandard optics from quantum space-time'',
Phys.\ Rev.\ D59 (1999) 124021
[{\underline{Uruguay}}] 413 citations.
\up

\noindent
[3] F.\ Pisano and V.\ Pleitez,
``An SU(3) x U(1) model for electroweak interactions'',
Phys.\ Rev.\ D46 (1992) 410
[{\underline{Brazil}}] 371 citations.
\up

\noindent
[4] J.\ Frenkel and J.C.\ Taylor,
``High Temperature Limit of Thermal QCD'',
Nucl.\ Phys.\ B334 (1990) 199
[{\underline{Brazil}}] 370 citations.
\up

\noindent
[5] N.\ Berkovits,
``Super-Poincar\'{e} covariant quantization of the superstring'',
J.\ High Energy Phys.\ 0004 (2000) 018
[{\underline{Brazil}}] 299 citations.
\up

\noindent
[6] L.P.\ Chimento, A.S.\ Jakubi, D.\ Pavon and W.\ Zimdahl,
``Interacting quin\-tessence solution to the coincidence problem'',
Phys.\ Rev.\ D67 (2003) 083513
[{\underline{Argentina}}] 263 citations.
\up

\noindent
[7] P.B.\ Arnold and O.\ Espinosa,
``Effective potential and first order phase transitions: 
Beyond leading-order'',
Phys.\ Rev.\ D47 (1993) 3546
[{\underline{Chile}}] 258 citations.
\up

\noindent
[8] G.\ Aldazabal, L.E.\ Ib\'{a}\~{n}ez, F.\ Quevedo and A.M.\ Uranga,
``D-branes at singularities: A bottom-up approach to the 
string embedding of the standard model'',
J.\ High Energy Phys.\ 0008 (2000) 002
[{\underline{Argentina}}] 254 citations. \\

Note that all authors of these top-cited papers are distinct.
This list is dominated by the nations of the leading group with
respect to the publication per capita, cf.\ Table \ref{pubpoptab}.

Five among these top-cited papers involve co-authors in
First World countries (there are no co-authors from other
countries in the Third World). References [3] and [5] are all
Brazilian, while in Ref.\ [1] all authors give a Chilean
institute as their first address, but part of them add a
second address in the First World.

\subsection{Groundbreaking earlier publications}

At last, we mention three groundbreaking highlights of earlier
theoretical HEP in LA, {\it i.e.}\ these three papers truly
entered the history of physics (cf.\ \cite{Masperi}):
the prediction of the geomagnetic effect on cosmic rays by
M.\ Sandoval Vallarta (MEX) and G.\ Lema\^{\i}tre \citep{Sandohist},
the prediction of the Z-boson by J.\ Leite Lopes (BRA) \citep{LLhist},
and the invention of dimensional regularization
by C.\ Bollini and J.J.\ Giambiagi (ARG) \citep{dimreghist}:\\

\noindent
G.\ Lema\^{\i}tre and M.\ Sandoval Vallarta, Phys.\ Rev.\ 43 (1933) 87. \\
J.\ Leite Lopes, Phys.\ Rev.\ 190 (1958) 509.\\
C.G.\ Bollini and J.J. Giambiagi, Nuovo Cim.\ 12B (1972) 20;
Phys.\ Lett.\ 40B (1972) 566.

\vspace*{-2mm}

\raggedright
\bibliography{references}

\end{document}